\documentclass[aps,prb,10pt,twocolumn,floatfix,superscriptaddress,showpacs,numerical]{revtex4-2}
\usepackage{amsmath, amssymb, braket}
\usepackage{graphicx}
\usepackage[T1]{fontenc}
\usepackage[utf8]{inputenc}
\usepackage{bbold}
\usepackage[colorlinks=true,linkcolor=blue,citecolor=blue,breaklinks]{hyperref}

\begin{document}
\title{Performance of the rigorous renormalization group for first order phase transitions and topological phases}
\author{Maxwell Block}
\thanks{M.~Block and J.~Motruk contributed equally to this work.}
\affiliation{Department of Physics, University of California, Berkeley, California 94720, USA}

\author{Johannes Motruk}
\thanks{M.~Block and J.~Motruk contributed equally to this work.}
\affiliation{Department of Physics, University of California, Berkeley, California 94720, USA}
\affiliation{Materials Sciences Division, Lawrence Berkeley National Laboratory, Berkeley, California 94720, USA}

\author{Snir Gazit}
\affiliation{Racah Institute of Physics and the Fritz Haber Center for Molecular Dynamics, The Hebrew University, Jerusalem 91904, Israel}

\author{Michael~P.~Zaletel}
\affiliation{Department of Physics, University of California, Berkeley, California 94720, USA}
\affiliation{Materials Sciences Division, Lawrence Berkeley National Laboratory, Berkeley, California 94720, USA}

\author{Zeph Landau}
\author{Umesh Vazirani}
\affiliation{Department of Electrical Engineering and Computer Sciences, University of California, Berkeley, California 94720, USA}

\author{Norman Y. Yao}
\affiliation{Department of Physics, University of California, Berkeley, California 94720, USA}
\affiliation{Materials Sciences Division, Lawrence Berkeley National Laboratory, Berkeley, California 94720, USA}

\date{\today}

\begin{abstract}
Expanding and improving the repertoire of numerical methods for studying quantum lattice models is an ongoing focus in many-body physics.
While the density matrix renormalization group (DMRG) has been established as a practically useful algorithm for finding the ground state in 1D systems, a provably efficient and accurate algorithm remained elusive until the introduction of the rigorous renormalization group (RRG) by Landau \textit{et al.} [\href{https://doi.org/10.1038/nphys3345}{Nature Physics \textbf{11}, 566 (2015)}].
In this paper, we study the accuracy and performance of a numerical implementation of RRG at first order phase transitions and in symmetry protected topological phases. 
Our study is motivated by the question of when RRG might provide a useful complement to the more established DMRG technique.
In particular, despite its general utility, DMRG can give unreliable results near first order phase transitions and in topological phases, since its local update procedure can fail to adequately explore (near) degenerate manifolds.
The rigorous theoretical underpinnings of RRG, meanwhile, suggest that it should not suffer from the same difficulties.
We show this optimism is justified, and that RRG indeed determines accurate, well-ordered energies even when DMRG does not.
Moreover, our performance analysis indicates that in certain circumstances \emph{seeding} DMRG with states determined by coarse runs of RRG may provide an advantage over simply performing DMRG.
\end{abstract}

\maketitle

\section{Introduction}

Developing numerical methods for determining the ground state of a quantum lattice model is a central and enduring challenge in condensed matter physics. 
For 1D systems, the density matrix renormalization group (DMRG)~\cite{White1992} has been established as an efficient and practical algorithm for studying large spin-chains and even quasi-2D systems~\cite{White1996,Stoudenmire2012,Zaletel2013,Motruk2016} and quantum chemistry problems~\cite{Chan2011,Wouters2014}.
In its modern formulation, DMRG is a variational algorithm in the space of matrix product states (MPS) that seeks to find the global energy minimum via local updates to the MPS ansatz~\cite{Schollwoeck2011}.
Empirically, DMRG succeeds in finding this minimum in the vast majority of cases for which it is employed. 
However, there is no proof that the algorithm converges to the true ground state. 
In order to fill this theoretical gap, a novel algorithm, termed the rigorous renormalization group (RRG) was recently introduced and proven to be able to find the ground state of gapped one-dimensional local Hamiltonians in a time polynomial in the system size~\cite{Landau2015}. 
 
There are a number of profound conceptual differences between RRG and DMRG, perhaps the most prominent being that RRG is \emph{not} variational.
Instead of relying on energy optimization, RRG makes use of an approximate ground-state projector (AGSP)~\cite{Arad2012}, an operator constructed from the Hamiltonian, which can be applied to an arbitrary subspace of states to increase its overlap with a span of low-energy states. 
Instead of attempting to determine a single low-energy eigenstate, RRG utilizes the AGSP to construct an entire low-energy \emph{subspace}.
RRG accomplishes this through a recursive procedure: first, the AGSP is applied to local approximations of the target subspace, increasing their accuracy, then these sub-spaces are merged to provide a larger-scale approximation \cite{Landau2015, Roberts2017}.
Remarkably, the properties of the AGSP make it possible to perform this merge in a manner which does not grow the subspace dimension but still improves the accuracy of the approximation. 
At its core, RRG embodies a renormalization procedure: local degrees of freedom are grouped together, then refined into a composite degree of freedom at a larger length scale.
 
In its original conception, the RRG algorithm contains certain steps that are challenging to implement numerically~\cite{Landau2015}; however, a slightly modified version has been studied by Roberts~\textit{et~al.}~\cite{Roberts2017}. 
Their work provides a proof of principle that RRG can be utilized as a numerical tool to compute low energy states of one-dimensional Hamiltonians.
Interestingly, for specifically-chosen models, such as the Bravyi-Gosset~\cite{Bravyi2015} and random XY model, RRG was even shown to outperform straightforward implementations of DMRG.
Modifying DMRG to use a sliding-block environment, however, restores its performance advantage for the Bravyi-Gosset case~\cite{Schmitteckert2018}.

With the utility of RRG established in these examples, one can ask a more general question: which situations in realistic physical systems, if any, might favor the use of RRG over DMRG?
An obvious weakness of DMRG is the fact that it may converge to a local energy minimum in the MPS manifold due to its variational nature and local optimization procedure~\cite{White1998,Dolfi2012}. 
This may constitute a problem in several cases
First, if there exist states that are close in energy, but very different in their global structure, DMRG may struggle to reliably find all states in the correct energetic order. 
This scenario arises naturally around first-order phase transitions~\cite{Motruk2017}.
Second, topological phases can exhibit large ground state degeneracies and, especially in quasi-2D applications of DMRG, it can be difficult to find the ground state in all topological sectors.
Third, long-range interactions require a ground state to satisfy non-local constraints, which poses a natural challenge for the local update procedure of DMRG.
Although approaches to improve the ergodicity of DMRG~\cite{White2005,Dolfi2012,Hubig2015} and to find degenerate topological states~\cite{He2014} have been put forward, there is still no guarantee that the algorithm converges to the state closest to the ground state in the MPS variational space.

In this work, we study the utility of RRG in the first two situations mentioned above: near first order phase transitions and in symmetry-protected topological (SPT) phases~\cite{Pollmann2010,Chen2011,Turner2011,Fidkowski2011,Schuch2011,SPTnote}.
Specifically, we compare the spectra determined by RRG and DMRG for a variety of models (Tab.~\ref{tab:models}) that span the full landscape of one dimensional spin chains.
As mentioned above, there are variety of implementations of DMRG; here, we compare to standard implementation provided by \cite{itensor}, which performs a two-site update on an initial random product state and includes density matrix corrections \cite{White2005}.
Our findings suggest that the numerical implementation of RRG may be suitable for studying 1D or quasi-1D models with competing orders and large degeneracy.  

The specific models we study are as follows.
For first order transitions, we examine three ferromagnetic models: the mixed field Ising model, the tricritical Ising model, and the anisotropic XXZ model.
These models allow us to compare RRG and DMRG performance across a rich variety of transitions: between ferromagnets, ferro- and para- magnets, and gapless and gapped phases. 
For SPT phases, we study two variants of the of the cluster model~\cite{Suzuki1971,Raussendorf2001}: one with additional antiferromagnetic coupling, and one with competing topological term.
These models allow us to expand our comparison to transitions between a Haldane chain~\cite{Haldane1983,Pollmann2010} and an antiferromagnet, and between a Haldane chain and a slightly less conventional SPT phase combined with symmetry breaking~\cite{Verresen2017}.

We now briefly summarize the organization of the paper. In, Sec.~\ref{sec:RRG}, we give a concise review the structure of the RRG algorithm. In Sec.~\ref{sec:results}, we present the results for the different models, and in Sec.~\ref{sec:perform} we evaluate the performance and resource needs of RRG. We end with a conclusion in Sec.~\ref{sec:conc}.

\section{The RRG Algorithm \label{sec:RRG}}

Here, we give a short review of the RRG algorithm.
For a more thorough discussion see Refs.~\cite{Landau2015} and \cite{Roberts2017}. 
The RRG algorithm relies crucially on an AGSP operator, $A$, which is roughly characterized by the property that it has the same eigenvectors  as the Hamiltonian under study, $H$, and additionally has small eigenvalues for excited states of $H$ (and hence approximately projects onto the low energy states). 
There are a variety of ways to construct the AGSP, but a numerically convenient technique is simply via imaginary time evolution~\footnote{In this work and in Ref.~\cite{Roberts2017}, a Suzuki-Trotter decomposition is used to efficiently compute an approximation to this operator.}:
\begin{align}
    A = \exp{\left[-\tau H \right]}.
\end{align} \label{eqn:AGSP-construction}
With the AGSP in hand we can now describe the main body of the algorithm.
The lattice is initially divided into $N = 2^m$ blocks. 
For each block $n$ we construct a local Hamiltonian $H_n$ consisting only of terms in $H$ that are supported entirely on that block.
We then diagonalize $H_n$ and keep the $s$ lowest eigenstates to form a local approximation to the low-energy subspace.

To make use of the AGSP, we decompose $A$ as a sum of terms which can individually be factored as a tensor product of terms acting to the left of $n$, on $n$, and to the right of $n$, i.e. $A = \sum_i \sum_j \lambda_i \mu_{ij} L^i_{<n} A^{ij}_n R^{ij}_{>n}$. 
This is accomplished via two Schmidt decompositions, and the $\lambda_i$ and $\mu_{ij}$ are the resulting Schmidt values.
We then take the $A^{ij}_n$ corresponding to the $D^2$ largest $\lambda_i \mu_{ij}$ and apply these local operators to the $s$ local low energy eigenstates.
This results in a new low-energy subspace supported on block $n$, now spanned by $sD^2$ states (some of which may be linearly dependent).

Finally, we merge the subspaces associated with blocks $n$ and $n+1$ ($n$ even). 
To do so, we first simply tensor product the states supported on blocks $n$, $n+1$ to form $s^2D^4$ states comprising a new two-block approximation for the low-energy subspace.
Then, we reduce the dimensionality of this subspace by diagonalizing the local Hamiltonian $H_{n, n+1}$ (i.e. consisting of terms from $H$ supported only on these blocks), restricted to the tensor product subspace.
After this diagonalization, we select the lowest $s$ energy eigenstates and use these as the new basis for the local approximation of the lower-energy subspace.

After completing this procedure for all pairs ($n$, $n+1$), we are left with $2^{m-1}$ blocks each associated with an $s$-dimensional low energy subspace.
Thus, we may repeat the application of the AGSP and merging procedure to generate successively larger approximations of the global low-energy subspace.
After $m$ steps, the process terminates and we are left with candidate eigenstates for the full system from the final diagonalization.

We note that our implementation of RRG, which is based on that of \cite{Roberts2017}, features two important differences form the version rigorously studied in \cite{Landau2015}.
First, rather than using a Chebyshev polynomial of the Hamiltonian to build the AGSP, we use Trotterized imaginary time evolution.
Second, instead of performing random sampling to reduce the dimensionality of merged subspaces we choose to keep a few low energy eigenstates of $H_{n, n+1}$.
The motivation for and consequences of these adjustments are discussed thoroughly in \cite{Roberts2017}.
Briefly, they make the algorithm easier to implement, and reduce constant factors in the runtime, at the cost of foregoing theoretical guarantees on accuracy \cite{Roberts2017}.

\begin{figure*}
    \includegraphics[scale=0.5]{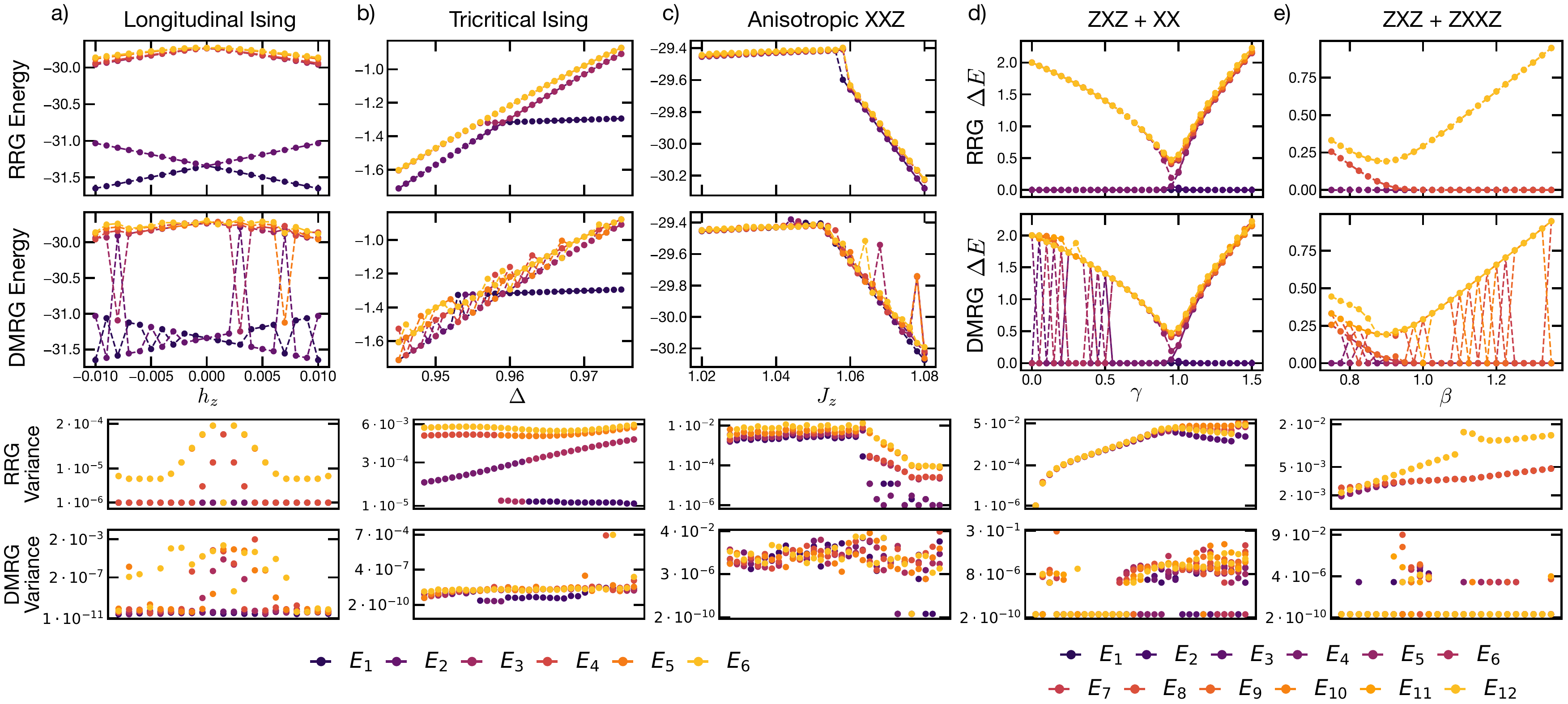}
    \caption{Comparison of energies and variances of the states found using RRG ($D=8$) and DMRG for the different models ($N=32$). (a)~Transverse field Ising model ($s$=6)~: DMRG finds states out of order and misses some low-energy states entirely while RRG computes the low-energy subspace reliably. (b)~Tricritical Ising model ($s=6$)~: For $\Delta$ just below the ferro-paramagnet transition, DMRG misses the ground state initially and converges to a paramagnetic state. (c)~Anisotropic $XXZ$ model ($s=6$)~: DMRG misses some states in the ferromagnetic phase; RRG is able to find the low-energy states in the gapless $XY$ up to to the critical point, where it has large error (d)~$ZXZ + XX$ model ($s=12$)~: DMRG struggles to produce the entire ground state manifold in the topological phase. (e)~$ZXZ+ZXXZ$ model ($s=12$)~: DMRG shows the same issues as in (d) in both topological phases. The energy variances of the DMRG states are very small in all models, mostly around the SVD cutoff ($\sim 10^{-10}$), except in some excited states and the gapless phase of the $XXZ$ model. RRG states display higher variances ($10^{-2} \sim 10^{-6}$) in all models.}
    \label{fig:main}
\end{figure*}

Finally, we remark that the RRG algorithm requires no variational optimization, and no specific ansatz for the form of the eigenstates. 
It is thus not susceptible to becoming stuck in sub-optimal minima, and does not exclude any subsets of Hilbert space by fiat.
Relatedly, one trades RRG accuracy for speed \emph{not} by further restricting a variational manifold but instead by adopting a coarser renormalization step.
Specifically, reducing $s$ and $D$ can vastly improve the speed (and reduce the memory needs) of RRG at the cost of finding fewer, and more inaccurate, states. 
We discuss this trade-off in more detail in Appendix \ref{app:hyper-params}, and focus the qualitative, generic features of RRG in the main text.
Generally, we hold $D$ fixed across the models with investigate but let $s$ vary between the models because of the large variability in ground state degeneracy (e.g., $s=6$ for the magnetic models while $s=12$ for Topological models in Fig.~\ref{fig:main}).

\renewcommand{\arraystretch}{1.5}
\begin{table*}
    \centering
    \begin{tabular}{c|c|c|c|c} 
        \hline \hline
        Model & Spin & Hamiltonian & Parameters & Properties \\
        \hline
        Longitudinal ising & $\frac 1 2$ & $-\sum \sigma^z \sigma^z + h_x \sigma^x + h_z \sigma^z$ & $h_x = 0.2$, $h_z=[-0.01,0.01]$ & F.O.P.T. ferro $\leftrightarrow$ ferro \\
        \hline
        Tricritcal Ising & 1 & $\sum -S^z S^z - h_x S^x + \Delta (S^z)^2$ & $h_x = 0.2$, $\Delta=[0.94,0.98]$ & F.O.P.T. ferro $\leftrightarrow$ para \\
        \hline
        Anisotropic XXZ & 1 & $\sum S^x S^x + S^y S^y - J_z S^z S^z + \Delta (S^z)^2$ & $\Delta = 0.1$, $J_z=[1.02,1.08]$ & F.O.P.T. ferro $\leftrightarrow$ gapless \\
        \hline
        ZXZ + ZXXZ & $ \frac 1 2 $ & $\sum \sigma^z \sigma^x \sigma^z + \beta \sigma^z \sigma^x \sigma^x \sigma^z + \gamma \sigma^x \sigma^x  + \delta \sigma^x$ & $\gamma = 0.2$, $\delta = 0.1$, $\beta=[0.75,1.35]$ & diff. topol. degeneracies \\
        \hline
        ZXZ + XX & $ \frac 1 2 $ & $\sum \sigma^z \sigma^x \sigma^z + \gamma \sigma^x \sigma^x$ & $\gamma = [0, 1.5]$ & topol. $\leftrightarrow$ anti-ferro\\
        \hline  \hline
    \end{tabular}
    \caption{Overview of the model Hamiltonians we consider including the parameter ranges we study. F.O.P.T. denotes a first order phase transition.}
    \label{tab:models}
\end{table*}

\section{Results \label{sec:results}}

\subsection{Mixed field Ising model}

Let us begin by exploring the performance of RRG in the context of a first order phase transition within the mixed field Ising model (Table~\ref{tab:models}):
\begin{equation}
 H_{\rm Ising} = - \sum_i \sigma^z_i \sigma^z_{i+1} + h_x \sigma_i^x + h_z \sigma_i^z.
\end{equation}
As demonstrated in Fig~\ref{fig:main}(a), for $|h_x| < 1$, this model undergoes a first order phase transition between a $\uparrow$-polarized ground-state and $\downarrow$-polarized ground state as the $h_z$ field is tuned.
Intuitively, near the transition, DMRG can become trapped locally optimizing a state of the incorrect polarization \footnote{As mentioned in the introduction, we use a DMRG implementation provided by \cite{itensor} that performs a 2-site update on an initial random product state and include density matrix corrections}.
A similar issue occurs for the excited states when DMRG seeks to optimize a magnon with the wrong polarization background.
Of course, the DMRG energies can always be sorted after they are determined, so the tendency of DMRG to find states out of order may appear to be only a mild inconvenience. 
However, this instability can lead to a more serious issue: sometimes DMRG will skip a low-energy state entirely and find an excited state instead.
In the mixed field Ising model, this is most clearly observed for $h_z \approx -0.01$ [Fig.~\ref{fig:main}(a)], where some low-lying excited states are missing from the DMRG spectrum entirely.

RRG, in contrast, consistently finds all the low energy eigenstates (Fig.~\ref{fig:main}).
This is not a trivial consequence of the fact that RRG targets a subspace (and hence cannot find energies out of order) --- rather, it demonstrates that the algorithm effectively combines locally non-optimal (energetically unfavorable) subspaces to build up a globally optimal (low-energy) subspace. 
To be more concrete, RRG first approximates the global low-energy subspace as a tensor product of local subspaces, each determined by diagonalizing a local Hamiltonian.
As long as each local subspace is at least two dimensional, it will include both $\uparrow$-polarized and $\downarrow$-polarized states.
Then, as these local subspaces are combined and refined, eigenstates from both polarization sectors can be included.

A first order transition driven by a symmetry breaking term inherently challenges DMRG, since it ensures the existence of nearly degenerate states that are not locally related. 
However, since the symmetry breaking is explicit in the Hamiltonian, DMRG can always be seeded with a state from the lower-energy broken-symmetry sector, e.g., with $\ket{\downarrow \cdots \downarrow}$ for $h_z > 0$ in this simple example.   
This motivates the examination of less contrived, and more interesting, models that challenge the local optimization of DMRG while also exhibiting ground-state order that cannot be straightforwardly read from the Hamiltonian.
%

\subsection{Tricritcal Ising model}

One such model is the tricritcal Ising model, described by the spin-1 hamiltonian
\begin{equation}
    H_{\rm tric} = \sum_i -S^z_i S^z_{i+1} - h_x S^z_i + \Delta (S^z_i)^2.
\end{equation}
Unlike the mixed field Ising model, the tricritical model exhibits a first order transition between fundamentally different phases: ferromagnetic ($\ket{m_s = \pm 1}$) at large $\Delta > 0$, and paramagnetic ($\ket{m_s=0}$) at small $\Delta > 0$ (for small $h_x$) [Fig.~\ref{fig:main}(b)].  
Since the onsite $(S^z)^2$ field always locally favors $\ket{m_s=0}$-aligned spins, DMRG is typically biased, at least in its first few sweeps, towards converging to the paramagnetic state. 
As a result, DMRG often fails to find the true ferromagnetic ground state near the transition, as evidenced by the overshooting ground state energy in Fig.~\ref{fig:main}.
Even deep in the ferromagnetic phase, DMRG often misses one of the pair of degenerate ground states, finding an $\ket{m_s=0}$ polarized state instead.
Crucially, there is no obvious way to seed DMRG to encourage it to find the correct ground state, without already knowing the location of the phase transition.
Similarly, without prior knowledge of the ground-state degeneracy, one must check the symmetry properties of the states found by DMRG to determine if it is missing representatives from the low-energy manifold.

On the other hand, since RRG initially determines and then merges local approximations of the low-energy subspace, it is not inherently biased towards constructing para- or ferromagnetic states.
Indeed, both para- and ferromagnetic states will be maintained in local subspaces, allowing RRG to reliably find the ground state(s) and the low-lying excited states [Fig.~\ref{fig:main}(b)].

\subsection{Anisotropic \texorpdfstring{$XXZ$}{XXZ} model}

We can compare the behavior of RRG and DMRG for another class of first order transitions through the anisotropic XXZ model, given by the spin-1 Hamiltonian
\begin{equation}
        H_{\rm xxz} = \sum_i S_i^x S_{i+1}^x + S_i^y S_{i+1}^y - J_z S_i^z S_{i+1}^z + \Delta (S_i^z)^2.
\end{equation}
This model exhibits a variety of phase transitions, but here we focus on the first order transition between a gapless $XY$ phase at small $J_z$ and ferromagnet at large $J_z$ (both at small, fixed $\Delta > 0$).
In the gapless phase, DMRG is able to robustly determine the low-energy spectrum using local updates, since long-wavelength excitations can be decomposed into a sequence of small local adjustments [Fig.~\ref{fig:main}(c)].
In the ferromagnetic phase, however, the inconsistency of DMRG re-surfaces, and it often fails to find both degenerate ground-states [Fig.~\ref{fig:main}(c)].
This difficulty can again be attributed to the onsite $(S^z)^2$, which biases local updates against ferromagnetic order.

On theoretical grounds, RRG is not expected to perform well in gapless phases, since it is difficult to efficiently approximate the AGSP in this setting \cite{arad_rigorous_2017}.
However, we observe that in practice RRG effectively determines the low-energy spectrum deep in the gapless $XY$ phase [Fig.~\ref{fig:main}(c)].
This is in contrast to the region near the transition, where RRG fails to find the ferromagnetic ground states for $J_z$ slightly larger than the critical value, resulting in a jump in the spectrum. [Fig.~\ref{fig:main}(c)].
The precise reason for this difficulty remains a topic for future research, but is presumably related to the complexity of approximating the AGSP at the critical point.
Deep in the ferromagnetic phase, RRG recovers and reliably finds all low energy eigenstates [Fig.~\ref{fig:main}(c)] as in the tricritical and mixed field Ising models.
This can again be attributed to the fact that RRG keeps the degenerate low-energy eigenstates on the local blocks which permit the construction of the full global low-energy space upon subsequent merging.

\subsection{\texorpdfstring{$ZXZ + XX$}{ZXZ+XX} model}

We now turn to the topological phases and first consider the cluster model with an additional antiferromagnetic term described by the Hamiltonian
\begin{equation}
 H_{\rm topo} = \sum_i \sigma^z_{i-1} \sigma^x_i \sigma^z_{i+1} + \gamma \sigma^x_i \sigma^x_{i+1}.
\end{equation}
The topological part of this Hamiltonian ($ZXZ$) can be understood through either the stabilizer formalism or via a Majorana representation~\cite{Verresen2017}.
From either perspective, it can be seen that the SPT ground-state factorizes into a short-range entangled bulk and two uncoupled edge spins as in the AKLT model~\cite{Affleck1987}.
This results in four-fold ground state degeneracy corresponding to the four possible edge-spin configurations: $\ket{\uparrow \uparrow}, \ket{\uparrow \downarrow}, \ket{\downarrow \uparrow}, \ket{\downarrow \downarrow}$ (\ref{fig:majo}).
As $\gamma$ is increased this system undergoes a transition from this topological phase to a antiferromagnet, with a concordant change to two-fold ground state degeneracy.
We note that in the topological phase, the entire ground-state subspace can be generated from a common short-range entangled bulk state plus local operations on edge spins.

Since there is no obstruction to changing the edge spin configuration during the DMRG optimization procedure, one would expect DMRG to consistently find the entire ground-state subspace in the topological phase.
We observe, however, that DMRG initialized with a random state often finds some excited states before all four ground-states, especially for $\gamma < 0.5$ as can be seen in Fig.~\ref{fig:main}(d). 
The reason for this behavior is that DMRG introduces a bulk excitation as it orthogonalizes the trial state against existing lower-energy states that have already been found.
This excitation is then never removed and steers the algorithm towards a local minimum, an excited state, to which it eventually converges.

For this model, the above problem can be conveniently addressed by cycling the DMRG initial conditions through all four edge-spin configurations.
However, this solution may not generalize to more complicated topological models, where the degenerate subspace has a more complex structure.

Interestingly, DMRG does find both degenerate ground states consistently on the antiferromagnetic side of the transition [Fig.~\ref{fig:main}(d)]; this is in contrast to the tricritical Ising and anisotropic $XXZ$ models, where it would often miss one of the states in the pair.
The differences behavior here can be explained by the fact that the $ZXZ + XX$ model is spin-$\frac 1 2$ and necessarily lacks the on-site $(\sigma^z)^2$ which term which biased DMRG against establishing ferromagnetic order.

RRG again demonstrates no instability it determining the eigenstates [Fig.~\ref{fig:main}] which reflects the fact that the cluster Hamiltonian is indeed a model that is particularly well suited for the algorithm.
Models of this kind are frustration free, and the exact ground state can be recovered by gluing together locally optimized subspaces meaning that the limit of $\gamma = 0$ would not even require the application of the AGSP.
Here, the correct ground state for a large system can be constructed from superpositions of tensor products out of the states in the ground state manifold of smaller blocks.
Our simulations demonstrate that this behavior extends to nonzero $\gamma$ and should therefore hold for general models of bosonic SPTs in one dimension.
%

\subsection{\texorpdfstring{$ZXZ+ZXXZ$}{ZXZ+ZXXZ} model}

We now consider a slightly more complicated topological model that features a transition between the SPT phase above and one with additional symmetry breaking: 
\begin{equation}
 \begin{split}
 H_{\rm topo} = &\sum_i \sigma^z \sigma^x \sigma^z + \beta \sigma^z \sigma^x \sigma^x \sigma^z \\
 & + \gamma \sigma^x \sigma^x  + \delta \sigma^x.
 \end{split}
\end{equation}
In particular, we consider small, fixed $\gamma$ and $\delta$ and vary $\beta \in [0.75, 1.35]$. 
At small $\beta$, the $ZXZ$ terms dominate and the system is in the AKLT phase; for $\beta \gtrsim 1.1$ the $ZXXZ$ terms gain the upper hand and the AKLT phase is replaced by a more complicated Majorana chain structure.
To gain some intuition for this new phase, consider the limiting case of $\beta \to \infty$.
In this limit, it is clear that the system decouples into three Majorana chains, with three Majorana modes at each edge.
These new modes pair up to form additional physical degree of freedoms, resulting in a ground-state degeneracy of $8$ (see Appendix~\ref{app:topo-models}).
Unlike in the AKLT phase, however, one of these pairings is \emph{non-local} --- Majoranas from opposite edges must be paired.
(This can also be easily understood form the stabilizer perspective, since $\pm ZZ \cdots ZZ$ is a stabilizer of the ground state.)
In the spin language, the additional factor two in the degeneracy is caused by symmetry breaking with order parameter $\sigma^z_{i-1} \sigma^y_i \sigma^z_{i+1}$ which is present on top of the SPT order~\cite{Verresen2017}.
%

Unlike the $ZXZ + XX$ model, this Hamiltonian features an on-site $X$ field that allows the local update procedure of DMRG to traverse all edge spin configurations.
Nonetheless, in \emph{both} topological phases, DMRG struggles to consistently find all degenerate ground states [Fig.~\ref{fig:main}(e)]. 
Moreover, in the $ZXXZ$ phase, cycling through edge-spin initial conditions no longer ensures DMRG will find the entire ground-state manifold, since there is a third delocalized qubit contributing to the degeneracy.
In principle, this can be overcome by initializing DMRG runs in different sectors of the broken symmetry in addition to the edge spin configurations, but the situation demonstrates that the initializing procedure can become rather involved in the presence of high ground state degeneracy of different physical origin.

RRG, on the other hand, consistently finds the complete spectrum without any fine-tuning of hyper-parameters or initial conditions [Fig.~\ref{fig:main}].
The results on this model demonstrate that RRG is also able to correctly determine the low-energy subspace when the ground state degeneracy is caused by a combination of topology and symmetry breaking.
%
%

\subsection{Seeding DMRG with RRG}

Since DMRG can be applied to arbitrary initial states, a natural question is whether it is possible to use RRG to determine good input states for DMRG.
This is especially relevant in light of the fact (discussed in more detail below) that current implementations of RRG are significantly slower than DMRG.
Therefore, one may hope to use a ``cheap'' initial run of RRG (i.e. with low $s$, $D$) to find approximate eigenstates that can then be further optimized by DMRG.
The upshot of this approach, compared to initializing DMRG with random states, is that the seeding procedure encourages DMRG to find all states of the low energy eigenspace without missing any, and in the correct order.

\begin{figure}
    \includegraphics[scale=0.49]{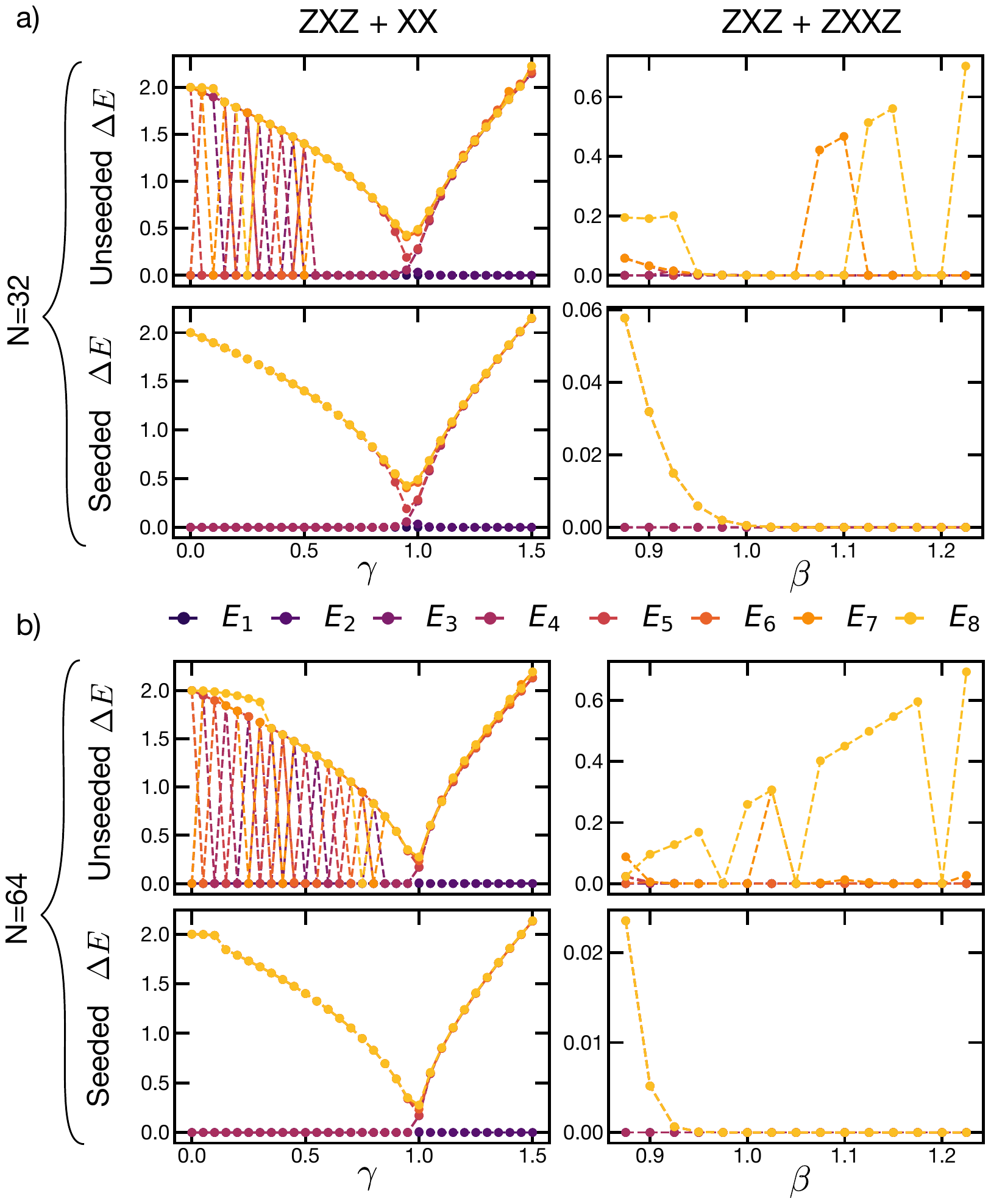}
    \caption{Comparison of DMRG energies given random initial statues (upper panels) and given RRG ($D=4$, $s=8$) initial states (lower panels). Seeding DMRG with states determined by RRG ameliorates the instability in energy order.}
    \label{fig:rrg-in}
\end{figure}

We investigate the utility of such a hybrid algorithm by applying it to the topological models ($ZXZ+ZXXZ$ and $ZXZ+XX$) discussed in the previous two subsections. 
We choose to focus on these models for two reasons: first, as observed above, the large degeneracy of their low energy manfolds makes it challenging for DMRG to find all ground-states consistently, making these models an ideal base case for \emph{improving} on DMRG; second, RRG performs especially well on these models (see Tab.~\ref{tab:resources} and Tab.~\ref{tab:app-resources}), making it plausible to use a single run of RRG in place of several trials of DMRG.
As suggested above, we first perform relatively inexpensive RRG ($D=4$), which is accurate enough to correctly determine degeneracies but still has significantly large energy errors (Fig.~\ref{fig:N32-D4}).
We then refine these candidate eigenstates with DMRG, increasing the accuracy of the energies.
Compared to DMRG with random initial states, the hybrid approach demonstrates significantly improved stability, and does not miss any degenerate states, see Fig.~\ref{fig:rrg-in}.
This advantage persists at larger systems size ($N=64$), where randomly initialized DMRG becomes even more unreliable.

It is also interesting to note that at larger system size, the RRG energies are very inaccurate, exhibiting an energy variance of $\sim 0.1$. 
After performing DMRG, however, the variances become similar to other DMRG runs ($\sim 10^{-4}$).
This behavior demonstrates that RRG need not be too energetically accurate to provide useful initial states for DMRG. 
It may therefore be possible to alter the RRG algorithm, or run it at exceedingly small $s$ and $D$, and still benefit from stabilizing DMRG when using its resulting states as input.
Indeed, a particularly promising application of RRG-seeded DMRG is for generating the manifold of degenerate states of an intrinsic topological phase using cylinder numerics~\cite{Jiang2012,Cincio2013,He2014,Szasz2020}.
Investigating this thoroughly for a range of models remains a topic for future research.

\renewcommand{\arraystretch}{1.5}
\begin{table*}
    \centering \begin{tabular}{c||c|c||c|c} 
        \hline 
        Model & \multicolumn{2}{c||}{RRG} & \multicolumn{2}{c}{DMRG} \\
        \hline \hline
         & Avg. Time [s] & Avg. Mem. [MB] & Avg. Time [s] & Avg. Mem. [MB] \\
        \hline
        Longitudinal Ising & $380$ & 68 & 4 & 0.50 \\
        \hline
        Tricritcal Ising & $13.5 \cdot 10^3$ & 570 & 24 & 0.76 \\        
        \hline
        Anisotropic XXZ & $45.9 \cdot 10^3$ & $2.0 \cdot 10^3$ & 590 & 4.1 \\
        \hline 
        ZXZ + XX & $25.9 \cdot 10^3$ & $1.6 \cdot 10^3$ & $3.7 \cdot 10^3$ & 9.9 \\
        \hline
        ZXZ + ZXXZ & $12.2 \cdot 10^3$ & $1.8 \cdot 10^3$ & $1.1 \cdot 10^3$ & 5.8 \\
        \hline
    \end{tabular}
    \caption{Computational resource comparison of RRG (D=8) and DMRG for the simulations shown in Fig.~\ref{fig:main}.}
    \label{tab:resources}
\end{table*}

\section{Performance \label{sec:perform}}

Although RRG offers an advantage over DMRG in stability and consistency, it comes at the cost of reduced performance in terms of accuracy, time and memory.
To characterize the accuracy of RRG and DMRG we consider the variances of the approximate eigenstates, defined as $\sigma_i^2 = \bra{\psi_i} H^2 \ket{\psi_i} - (\bra{\psi_i} H \ket{\psi_i})^2$. 
A true eigenstate would have zero variance, and large variance indicates an inaccurate approximation.
The variances of the DMRG ground states are typically on the order of the SVD truncation error ($\sim 10^{-10}$), but can be larger --- particularly for excited states and in the gapless phase of the anisotropic XXZ model (Fig.~\ref{fig:main}). 
The variances of the RRG states range from $10^{-6} - 10^{-2}$, and vary significantly between models and even between phases within a model. 
This variation is driven by how accurately the AGSP is approximated globally by the Suzuki-Trotter decomposition, and locally by the $D^2$ largest Schmidt value operators. 
For some applications, the variances of the RRG states may be acceptable.
Otherwise, the RRG states can be used as input for further optimization by DMRG.

As can be seen in the resource summary table, Tab.~\ref{tab:resources}, RRG requires significantly more time and memory than DMRG to analyze the same models. 
However, setting large gap in performance aside, the RRG resource requirements themselves provide insight into the strengths and weaknesses of the algorithm.
For example, RRG is slowest and most memory intensive for the XXZ model, which is naturally explained by the fact the its rigorous antecedent has worse performance for gapless systems \cite{arad_rigorous_2017}.
Conversely, RRG performs relatively well on the topological models, and is only $10 \times$ slower than DMRG for the $ZXZ + XX$ (or $5 \times$ slower for the $D=4$ RRG runs; see Tab.~\ref{tab:app-resources}).
This is a consequence of the fact that the RRG merging procedure is optimal for frustration free hamiltonians, and the advantage evidently carries over to only approximately frustration free models, or models belonging to the same phase as a frustration free fixed point.

\section{Conclusion \label{sec:conc}}

Although DMRG is a powerful and flexible algorithm for studying 1D quantum systems, it still has difficulty recovering the low-energy spectra of models with degeneracy. 
For simple models, this shortcoming can be overcome by careful initialization based on physical intuition.
In more complex systems, however, such an \emph{a priori} understanding may not be possible.
In these scenarios, RRG is an ideal tool due to its reliability in finding low energy eigenstates.
We have demonstrated this reliability in systems featuring first-order transitions and topological phases.
Moreover, the theoretical underpinning of the algorithm suggests that this stability extends to many 1D systems of interest.
For example, while the proof of convergence~\cite{Landau2015} and first numerical test~\cite{Roberts2017} of RRG were based on Hamiltonians with nearest-neighbor interactions, the success of the method in the topological models demonstrates its applicability to systems with longer range couplings.

In accordance with Ref.~\cite{Roberts2017}, we find that the main drawback of RRG consists of the long runtimes.
There are several directions in which implementations of the algorithm can be optimized.
First, RRG performance --- in terms of both accuracy and resources --- is ultimately limited by how faithfully the AGSP can be represented as an MPO of tractable dimension.
It may therefore be possible to improve the numerical implementation of RRG by adopting more sophisticated routines for approximating the AGSP, such as those outlined in the original convergence proof.
This represents an interesting future research direction for AGSPs and their tensor network representation in general.
Second, the RRG algorithm hosts great potential for parallelization. 
Within a tree level, the expansion of the subspace in one merge step and the computation of the effective Hamiltonian on the restricted Hilbert space may easily be performed simultaneously for all pairs of blocks.
Furthermore, even the expansion and Hamiltonian generation within one block may be parallelized as a set of states is independently expanded by the application of a set of operators.
This is in contrast to DMRG where parallelization is currently only possible to a limited extent~\cite{Stoudenmire2013,Levy2020}, yet distribution of workload on multiple nodes is key in modern supercomputer structures.
Finally, Krylov subspace methods can be straightforwardly applied to diagonalizing restricted Hamiltonians~\cite{RRGcode}.

More broadly, it would be interesting to implement the RRG algorithm on non-MPS wavefunction ans\"atze, and to explore other operator representations of the AGSP.
For example, neural network states, especially restricted Boltzmann machines (RBM)~\cite{Smolensky1986} have been suggested as a promising variational ansatz to represent quantum many-body states~\cite{Carleo2017}.
One could adapt RRG to this representation and compare its performance to other methods of optimizing RBMs.

\section*{Acknowldegements}

The calculations in this work have been performed with an extension of the code provided in~\cite{RRGcode}.
The authors thank Laurens Vanderstraeten for useful discussions.
J.M.~received funding by TIMES at Lawrence Berkeley National Laboratory supported by the U. S. Department of Energy, Office of Basic Energy Sciences, Division of Materials Sciences and Engineering, under Contract No. DE-AC02-76SF00515 and by DFG research fellowship No. MO 3278/1-1.
M.B.~acknowledges support through the Department of Defense (DoD) through the National Defense Science \& Engineering Graduate (NDSEG) Fellowship Program.
MPZ was funded by the U.S. Department of Energy, Office of Science, Office of Basic Energy Sciences, Materials Sciences and Engineering Division under Contract No. DE-AC02-05- CH11231 through the Scientific Discovery through Advanced Computing (SciDAC) program (KC23DAC Topological and Correlated Matter via Tensor Networks and Quantum Monte Carlo).
S.G. acknowledges support from the Israel Science Foundation, Grant No. 1686/18.

\bibliography{rrg.bib}

\begin{thebibliography}{42}%
\makeatletter
\providecommand \@ifxundefined [1]{%
 \@ifx{#1\undefined}
}%
\providecommand \@ifnum [1]{%
 \ifnum #1\expandafter \@firstoftwo
 \else \expandafter \@secondoftwo
 \fi
}%
\providecommand \@ifx [1]{%
 \ifx #1\expandafter \@firstoftwo
 \else \expandafter \@secondoftwo
 \fi
}%
\providecommand \natexlab [1]{#1}%
\providecommand \enquote  [1]{``#1''}%
\providecommand \bibnamefont  [1]{#1}%
\providecommand \bibfnamefont [1]{#1}%
\providecommand \citenamefont [1]{#1}%
\providecommand \href@noop [0]{\@secondoftwo}%
\providecommand \href [0]{\begingroup \@sanitize@url \@href}%
\providecommand \@href[1]{\@@startlink{#1}\@@href}%
\providecommand \@@href[1]{\endgroup#1\@@endlink}%
\providecommand \@sanitize@url [0]{\catcode `\\12\catcode `\$12\catcode
  `\&12\catcode `\#12\catcode `\^12\catcode `\_12\catcode `\%12\relax}%
\providecommand \@@startlink[1]{}%
\providecommand \@@endlink[0]{}%
\providecommand \url  [0]{\begingroup\@sanitize@url \@url }%
\providecommand \@url [1]{\endgroup\@href {#1}{\urlprefix }}%
\providecommand \urlprefix  [0]{URL }%
\providecommand \Eprint [0]{\href }%
\providecommand \doibase [0]{https://doi.org/}%
\providecommand \selectlanguage [0]{\@gobble}%
\providecommand \bibinfo  [0]{\@secondoftwo}%
\providecommand \bibfield  [0]{\@secondoftwo}%
\providecommand \translation [1]{[#1]}%
\providecommand \BibitemOpen [0]{}%
\providecommand \bibitemStop [0]{}%
\providecommand \bibitemNoStop [0]{.\EOS\space}%
\providecommand \EOS [0]{\spacefactor3000\relax}%
\providecommand \BibitemShut  [1]{\csname bibitem#1\endcsname}%
\let\auto@bib@innerbib\@empty
\bibitem [{\citenamefont {White}(1992)}]{White1992}%
  \BibitemOpen
  \bibfield  {author} {\bibinfo {author} {\bibfnamefont {S.~R.}\ \bibnamefont
  {White}},\ }\bibfield  {title} {\bibinfo {title} {Density matrix formulation
  for quantum renormalization groups},\ }\href
  {https://doi.org/10.1103/PhysRevLett.69.2863} {\bibfield  {journal} {\bibinfo
   {journal} {Phys. Rev. Lett.}\ }\textbf {\bibinfo {volume} {69}},\ \bibinfo
  {pages} {2863} (\bibinfo {year} {1992})}\BibitemShut {NoStop}%
\bibitem [{\citenamefont {White}(1996)}]{White1996}%
  \BibitemOpen
  \bibfield  {author} {\bibinfo {author} {\bibfnamefont {S.~R.}\ \bibnamefont
  {White}},\ }\bibfield  {title} {\bibinfo {title} {{Spin Gaps in a Frustrated
  Heisenberg Model for ${\mathrm{CaV}}_{4}\mathrm{O}_{9}$}},\ }\href
  {https://doi.org/10.1103/PhysRevLett.77.3633} {\bibfield  {journal} {\bibinfo
   {journal} {Phys. Rev. Lett.}\ }\textbf {\bibinfo {volume} {77}},\ \bibinfo
  {pages} {3633} (\bibinfo {year} {1996})}\BibitemShut {NoStop}%
\bibitem [{\citenamefont {Stoudenmire}\ and\ \citenamefont
  {White}(2012)}]{Stoudenmire2012}%
  \BibitemOpen
  \bibfield  {author} {\bibinfo {author} {\bibfnamefont {E.}~\bibnamefont
  {Stoudenmire}}\ and\ \bibinfo {author} {\bibfnamefont {S.~R.}\ \bibnamefont
  {White}},\ }\bibfield  {title} {\bibinfo {title} {{Studying Two-Dimensional
  Systems with the Density Matrix Renormalization Group}},\ }\href
  {https://doi.org/10.1146/annurev-conmatphys-020911-125018} {\bibfield
  {journal} {\bibinfo  {journal} {Annual Review of Condensed Matter Physics}\
  }\textbf {\bibinfo {volume} {3}},\ \bibinfo {pages} {111} (\bibinfo {year}
  {2012})}\BibitemShut {NoStop}%
\bibitem [{\citenamefont {Zaletel}\ \emph {et~al.}(2013)\citenamefont
  {Zaletel}, \citenamefont {Mong},\ and\ \citenamefont
  {Pollmann}}]{Zaletel2013}%
  \BibitemOpen
  \bibfield  {author} {\bibinfo {author} {\bibfnamefont {M.~P.}\ \bibnamefont
  {Zaletel}}, \bibinfo {author} {\bibfnamefont {R.~S.~K.}\ \bibnamefont
  {Mong}},\ and\ \bibinfo {author} {\bibfnamefont {F.}~\bibnamefont
  {Pollmann}},\ }\bibfield  {title} {\bibinfo {title} {{Topological
  Characterization of Fractional Quantum Hall Ground States from Microscopic
  Hamiltonians}},\ }\href {https://doi.org/10.1103/PhysRevLett.110.236801}
  {\bibfield  {journal} {\bibinfo  {journal} {Phys. Rev. Lett.}\ }\textbf
  {\bibinfo {volume} {110}},\ \bibinfo {pages} {236801} (\bibinfo {year}
  {2013})}\BibitemShut {NoStop}%
\bibitem [{\citenamefont {Motruk}\ \emph {et~al.}(2016)\citenamefont {Motruk},
  \citenamefont {Zaletel}, \citenamefont {Mong},\ and\ \citenamefont
  {Pollmann}}]{Motruk2016}%
  \BibitemOpen
  \bibfield  {author} {\bibinfo {author} {\bibfnamefont {J.}~\bibnamefont
  {Motruk}}, \bibinfo {author} {\bibfnamefont {M.~P.}\ \bibnamefont {Zaletel}},
  \bibinfo {author} {\bibfnamefont {R.~S.~K.}\ \bibnamefont {Mong}},\ and\
  \bibinfo {author} {\bibfnamefont {F.}~\bibnamefont {Pollmann}},\ }\bibfield
  {title} {\bibinfo {title} {Density matrix renormalization group on a cylinder
  in mixed real and momentum space},\ }\href
  {https://doi.org/10.1103/PhysRevB.93.155139} {\bibfield  {journal} {\bibinfo
  {journal} {Phys. Rev. B}\ }\textbf {\bibinfo {volume} {93}},\ \bibinfo
  {pages} {155139} (\bibinfo {year} {2016})}\BibitemShut {NoStop}%
\bibitem [{\citenamefont {Chan}\ and\ \citenamefont {Sharma}(2011)}]{Chan2011}%
  \BibitemOpen
  \bibfield  {author} {\bibinfo {author} {\bibfnamefont {G.~K.-L.}\
  \bibnamefont {Chan}}\ and\ \bibinfo {author} {\bibfnamefont {S.}~\bibnamefont
  {Sharma}},\ }\bibfield  {title} {\bibinfo {title} {The density matrix
  renormalization group in quantum chemistry},\ }\href
  {https://doi.org/10.1146/annurev-physchem-032210-103338} {\bibfield
  {journal} {\bibinfo  {journal} {Annual Review of Physical Chemistry}\
  }\textbf {\bibinfo {volume} {62}},\ \bibinfo {pages} {465} (\bibinfo {year}
  {2011})}\BibitemShut {NoStop}%
\bibitem [{\citenamefont {Wouters}\ and\ \citenamefont
  {Van~Neck}(2014)}]{Wouters2014}%
  \BibitemOpen
  \bibfield  {author} {\bibinfo {author} {\bibfnamefont {S.}~\bibnamefont
  {Wouters}}\ and\ \bibinfo {author} {\bibfnamefont {D.}~\bibnamefont
  {Van~Neck}},\ }\bibfield  {title} {\bibinfo {title} {The density matrix
  renormalization group for ab initio quantum chemistry},\ }\href
  {https://doi.org/10.1140/epjd/e2014-50500-1} {\bibfield  {journal} {\bibinfo
  {journal} {The European Physical Journal D}\ }\textbf {\bibinfo {volume}
  {68}},\ \bibinfo {pages} {272} (\bibinfo {year} {2014})}\BibitemShut
  {NoStop}%
\bibitem [{\citenamefont {Schollwöck}(2011)}]{Schollwoeck2011}%
  \BibitemOpen
  \bibfield  {author} {\bibinfo {author} {\bibfnamefont {U.}~\bibnamefont
  {Schollwöck}},\ }\bibfield  {title} {\bibinfo {title} {The density-matrix
  renormalization group in the age of matrix product states},\ }\href
  {https://doi.org/http://dx.doi.org/10.1016/j.aop.2010.09.012} {\bibfield
  {journal} {\bibinfo  {journal} {Annals of Physics}\ }\textbf {\bibinfo
  {volume} {326}},\ \bibinfo {pages} {96 } (\bibinfo {year}
  {2011})}\BibitemShut {NoStop}%
\bibitem [{\citenamefont {Landau}\ \emph {et~al.}(2015)\citenamefont {Landau},
  \citenamefont {Vazirani},\ and\ \citenamefont {Vidick}}]{Landau2015}%
  \BibitemOpen
  \bibfield  {author} {\bibinfo {author} {\bibfnamefont {Z.}~\bibnamefont
  {Landau}}, \bibinfo {author} {\bibfnamefont {U.}~\bibnamefont {Vazirani}},\
  and\ \bibinfo {author} {\bibfnamefont {T.}~\bibnamefont {Vidick}},\
  }\bibfield  {title} {\bibinfo {title} {A polynomial time algorithm for the
  ground state of one-dimensional gapped local {Hamiltonians}},\ }\href
  {https://doi.org/10.1038/nphys3345} {\bibfield  {journal} {\bibinfo
  {journal} {Nature Physics}\ }\textbf {\bibinfo {volume} {11}},\ \bibinfo
  {pages} {566} (\bibinfo {year} {2015})}\BibitemShut {NoStop}%
\bibitem [{\citenamefont {Arad}\ \emph {et~al.}(2012)\citenamefont {Arad},
  \citenamefont {Landau},\ and\ \citenamefont {Vazirani}}]{Arad2012}%
  \BibitemOpen
  \bibfield  {author} {\bibinfo {author} {\bibfnamefont {I.}~\bibnamefont
  {Arad}}, \bibinfo {author} {\bibfnamefont {Z.}~\bibnamefont {Landau}},\ and\
  \bibinfo {author} {\bibfnamefont {U.}~\bibnamefont {Vazirani}},\ }\bibfield
  {title} {\bibinfo {title} {Improved one-dimensional area law for
  frustration-free systems},\ }\href
  {https://doi.org/10.1103/PhysRevB.85.195145} {\bibfield  {journal} {\bibinfo
  {journal} {Phys. Rev. B}\ }\textbf {\bibinfo {volume} {85}},\ \bibinfo
  {pages} {195145} (\bibinfo {year} {2012})}\BibitemShut {NoStop}%
\bibitem [{\citenamefont {Roberts}\ \emph {et~al.}(2017)\citenamefont
  {Roberts}, \citenamefont {Vidick},\ and\ \citenamefont
  {Motrunich}}]{Roberts2017}%
  \BibitemOpen
  \bibfield  {author} {\bibinfo {author} {\bibfnamefont {B.}~\bibnamefont
  {Roberts}}, \bibinfo {author} {\bibfnamefont {T.}~\bibnamefont {Vidick}},\
  and\ \bibinfo {author} {\bibfnamefont {O.~I.}\ \bibnamefont {Motrunich}},\
  }\bibfield  {title} {\bibinfo {title} {Implementation of rigorous
  renormalization group method for ground space and low-energy states of local
  {Hamiltonians}},\ }\href {https://doi.org/10.1103/PhysRevB.96.214203}
  {\bibfield  {journal} {\bibinfo  {journal} {Phys. Rev. B}\ }\textbf {\bibinfo
  {volume} {96}},\ \bibinfo {pages} {214203} (\bibinfo {year}
  {2017})}\BibitemShut {NoStop}%
\bibitem [{\citenamefont {Bravyi}\ and\ \citenamefont
  {Gosset}(2015)}]{Bravyi2015}%
  \BibitemOpen
  \bibfield  {author} {\bibinfo {author} {\bibfnamefont {S.}~\bibnamefont
  {Bravyi}}\ and\ \bibinfo {author} {\bibfnamefont {D.}~\bibnamefont
  {Gosset}},\ }\bibfield  {title} {\bibinfo {title} {Gapped and gapless phases
  of frustration-free spin-12 chains},\ }\href
  {https://doi.org/10.1063/1.4922508} {\bibfield  {journal} {\bibinfo
  {journal} {Journal of Mathematical Physics}\ }\textbf {\bibinfo {volume}
  {56}},\ \bibinfo {pages} {061902} (\bibinfo {year} {2015})}\BibitemShut
  {NoStop}%
\bibitem [{\citenamefont {Schmitteckert}(2018)}]{Schmitteckert2018}%
  \BibitemOpen
  \bibfield  {author} {\bibinfo {author} {\bibfnamefont {P.}~\bibnamefont
  {Schmitteckert}},\ }\bibfield  {title} {\bibinfo {title} {Density matrix
  renormalization group for a highly degenerate quantum system: Sliding
  environment block approach},\ }\href
  {https://doi.org/10.1103/PhysRevB.97.161110} {\bibfield  {journal} {\bibinfo
  {journal} {Phys. Rev. B}\ }\textbf {\bibinfo {volume} {97}},\ \bibinfo
  {pages} {161110(R)} (\bibinfo {year} {2018})}\BibitemShut {NoStop}%
\bibitem [{\citenamefont {White}\ and\ \citenamefont
  {Scalapino}(1998)}]{White1998}%
  \BibitemOpen
  \bibfield  {author} {\bibinfo {author} {\bibfnamefont {S.~R.}\ \bibnamefont
  {White}}\ and\ \bibinfo {author} {\bibfnamefont {D.~J.}\ \bibnamefont
  {Scalapino}},\ }\bibfield  {title} {\bibinfo {title} {{Density Matrix
  Renormalization Group Study of the Striped Phase in the 2D
  $\mathit{t}\ensuremath{-}\mathit{J}$ Model}},\ }\href
  {https://doi.org/10.1103/PhysRevLett.80.1272} {\bibfield  {journal} {\bibinfo
   {journal} {Phys. Rev. Lett.}\ }\textbf {\bibinfo {volume} {80}},\ \bibinfo
  {pages} {1272} (\bibinfo {year} {1998})}\BibitemShut {NoStop}%
\bibitem [{\citenamefont {Dolfi}\ \emph {et~al.}(2012)\citenamefont {Dolfi},
  \citenamefont {Bauer}, \citenamefont {Troyer},\ and\ \citenamefont
  {Ristivojevic}}]{Dolfi2012}%
  \BibitemOpen
  \bibfield  {author} {\bibinfo {author} {\bibfnamefont {M.}~\bibnamefont
  {Dolfi}}, \bibinfo {author} {\bibfnamefont {B.}~\bibnamefont {Bauer}},
  \bibinfo {author} {\bibfnamefont {M.}~\bibnamefont {Troyer}},\ and\ \bibinfo
  {author} {\bibfnamefont {Z.}~\bibnamefont {Ristivojevic}},\ }\bibfield
  {title} {\bibinfo {title} {{Multigrid Algorithms for Tensor Network
  States}},\ }\href {https://doi.org/10.1103/PhysRevLett.109.020604} {\bibfield
   {journal} {\bibinfo  {journal} {Phys. Rev. Lett.}\ }\textbf {\bibinfo
  {volume} {109}},\ \bibinfo {pages} {020604} (\bibinfo {year}
  {2012})}\BibitemShut {NoStop}%
\bibitem [{\citenamefont {Motruk}\ and\ \citenamefont
  {Pollmann}(2017)}]{Motruk2017}%
  \BibitemOpen
  \bibfield  {author} {\bibinfo {author} {\bibfnamefont {J.}~\bibnamefont
  {Motruk}}\ and\ \bibinfo {author} {\bibfnamefont {F.}~\bibnamefont
  {Pollmann}},\ }\bibfield  {title} {\bibinfo {title} {Phase transitions and
  adiabatic preparation of a fractional {Chern} insulator in a boson cold-atom
  model},\ }\href {https://doi.org/10.1103/PhysRevB.96.165107} {\bibfield
  {journal} {\bibinfo  {journal} {Phys. Rev. B}\ }\textbf {\bibinfo {volume}
  {96}},\ \bibinfo {pages} {165107} (\bibinfo {year} {2017})}\BibitemShut
  {NoStop}%
\bibitem [{\citenamefont {White}(2005)}]{White2005}%
  \BibitemOpen
  \bibfield  {author} {\bibinfo {author} {\bibfnamefont {S.~R.}\ \bibnamefont
  {White}},\ }\bibfield  {title} {\bibinfo {title} {Density matrix
  renormalization group algorithms with a single center site},\ }\href
  {https://doi.org/10.1103/PhysRevB.72.180403} {\bibfield  {journal} {\bibinfo
  {journal} {Phys. Rev. B}\ }\textbf {\bibinfo {volume} {72}},\ \bibinfo
  {pages} {180403(R)} (\bibinfo {year} {2005})}\BibitemShut {NoStop}%
\bibitem [{\citenamefont {Hubig}\ \emph {et~al.}(2015)\citenamefont {Hubig},
  \citenamefont {McCulloch}, \citenamefont {Schollw\"ock},\ and\ \citenamefont
  {Wolf}}]{Hubig2015}%
  \BibitemOpen
  \bibfield  {author} {\bibinfo {author} {\bibfnamefont {C.}~\bibnamefont
  {Hubig}}, \bibinfo {author} {\bibfnamefont {I.~P.}\ \bibnamefont
  {McCulloch}}, \bibinfo {author} {\bibfnamefont {U.}~\bibnamefont
  {Schollw\"ock}},\ and\ \bibinfo {author} {\bibfnamefont {F.~A.}\ \bibnamefont
  {Wolf}},\ }\bibfield  {title} {\bibinfo {title} {Strictly single-site dmrg
  algorithm with subspace expansion},\ }\href
  {https://doi.org/10.1103/PhysRevB.91.155115} {\bibfield  {journal} {\bibinfo
  {journal} {Phys. Rev. B}\ }\textbf {\bibinfo {volume} {91}},\ \bibinfo
  {pages} {155115} (\bibinfo {year} {2015})}\BibitemShut {NoStop}%
\bibitem [{\citenamefont {He}\ \emph {et~al.}(2014)\citenamefont {He},
  \citenamefont {Sheng},\ and\ \citenamefont {Chen}}]{He2014}%
  \BibitemOpen
  \bibfield  {author} {\bibinfo {author} {\bibfnamefont {Y.-C.}\ \bibnamefont
  {He}}, \bibinfo {author} {\bibfnamefont {D.~N.}\ \bibnamefont {Sheng}},\ and\
  \bibinfo {author} {\bibfnamefont {Y.}~\bibnamefont {Chen}},\ }\bibfield
  {title} {\bibinfo {title} {Obtaining topological degenerate ground states by
  the density matrix renormalization group},\ }\href
  {https://doi.org/10.1103/PhysRevB.89.075110} {\bibfield  {journal} {\bibinfo
  {journal} {Phys. Rev. B}\ }\textbf {\bibinfo {volume} {89}},\ \bibinfo
  {pages} {075110} (\bibinfo {year} {2014})}\BibitemShut {NoStop}%
\bibitem [{\citenamefont {Pollmann}\ \emph {et~al.}(2010)\citenamefont
  {Pollmann}, \citenamefont {Turner}, \citenamefont {Berg},\ and\ \citenamefont
  {Oshikawa}}]{Pollmann2010}%
  \BibitemOpen
  \bibfield  {author} {\bibinfo {author} {\bibfnamefont {F.}~\bibnamefont
  {Pollmann}}, \bibinfo {author} {\bibfnamefont {A.~M.}\ \bibnamefont
  {Turner}}, \bibinfo {author} {\bibfnamefont {E.}~\bibnamefont {Berg}},\ and\
  \bibinfo {author} {\bibfnamefont {M.}~\bibnamefont {Oshikawa}},\ }\bibfield
  {title} {\bibinfo {title} {Entanglement spectrum of a topological phase in
  one dimension},\ }\href {https://doi.org/10.1103/PhysRevB.81.064439}
  {\bibfield  {journal} {\bibinfo  {journal} {Phys. Rev. B}\ }\textbf {\bibinfo
  {volume} {81}},\ \bibinfo {pages} {064439} (\bibinfo {year}
  {2010})}\BibitemShut {NoStop}%
\bibitem [{\citenamefont {Chen}\ \emph {et~al.}(2011)\citenamefont {Chen},
  \citenamefont {Gu},\ and\ \citenamefont {Wen}}]{Chen2011}%
  \BibitemOpen
  \bibfield  {author} {\bibinfo {author} {\bibfnamefont {X.}~\bibnamefont
  {Chen}}, \bibinfo {author} {\bibfnamefont {Z.-C.}\ \bibnamefont {Gu}},\ and\
  \bibinfo {author} {\bibfnamefont {X.-G.}\ \bibnamefont {Wen}},\ }\bibfield
  {title} {\bibinfo {title} {Classification of gapped symmetric phases in
  one-dimensional spin systems},\ }\href
  {https://doi.org/10.1103/PhysRevB.83.035107} {\bibfield  {journal} {\bibinfo
  {journal} {Phys. Rev. B}\ }\textbf {\bibinfo {volume} {83}},\ \bibinfo
  {pages} {035107} (\bibinfo {year} {2011})}\BibitemShut {NoStop}%
\bibitem [{\citenamefont {Turner}\ \emph {et~al.}(2011)\citenamefont {Turner},
  \citenamefont {Pollmann},\ and\ \citenamefont {Berg}}]{Turner2011}%
  \BibitemOpen
  \bibfield  {author} {\bibinfo {author} {\bibfnamefont {A.~M.}\ \bibnamefont
  {Turner}}, \bibinfo {author} {\bibfnamefont {F.}~\bibnamefont {Pollmann}},\
  and\ \bibinfo {author} {\bibfnamefont {E.}~\bibnamefont {Berg}},\ }\bibfield
  {title} {\bibinfo {title} {Topological phases of one-dimensional fermions: An
  entanglement point of view},\ }\href
  {https://doi.org/10.1103/PhysRevB.83.075102} {\bibfield  {journal} {\bibinfo
  {journal} {Phys. Rev. B}\ }\textbf {\bibinfo {volume} {83}},\ \bibinfo
  {pages} {075102} (\bibinfo {year} {2011})}\BibitemShut {NoStop}%
\bibitem [{\citenamefont {Fidkowski}\ and\ \citenamefont
  {Kitaev}(2011)}]{Fidkowski2011}%
  \BibitemOpen
  \bibfield  {author} {\bibinfo {author} {\bibfnamefont {L.}~\bibnamefont
  {Fidkowski}}\ and\ \bibinfo {author} {\bibfnamefont {A.}~\bibnamefont
  {Kitaev}},\ }\bibfield  {title} {\bibinfo {title} {Topological phases of
  fermions in one dimension},\ }\href
  {https://doi.org/10.1103/PhysRevB.83.075103} {\bibfield  {journal} {\bibinfo
  {journal} {Phys. Rev. B}\ }\textbf {\bibinfo {volume} {83}},\ \bibinfo
  {pages} {075103} (\bibinfo {year} {2011})}\BibitemShut {NoStop}%
\bibitem [{\citenamefont {Schuch}\ \emph {et~al.}(2011)\citenamefont {Schuch},
  \citenamefont {P\'erez-Garc\'{\i}a},\ and\ \citenamefont
  {Cirac}}]{Schuch2011}%
  \BibitemOpen
  \bibfield  {author} {\bibinfo {author} {\bibfnamefont {N.}~\bibnamefont
  {Schuch}}, \bibinfo {author} {\bibfnamefont {D.}~\bibnamefont
  {P\'erez-Garc\'{\i}a}},\ and\ \bibinfo {author} {\bibfnamefont
  {I.}~\bibnamefont {Cirac}},\ }\bibfield  {title} {\bibinfo {title}
  {Classifying quantum phases using matrix product states and projected
  entangled pair states},\ }\href {https://doi.org/10.1103/PhysRevB.84.165139}
  {\bibfield  {journal} {\bibinfo  {journal} {Phys. Rev. B}\ }\textbf {\bibinfo
  {volume} {84}},\ \bibinfo {pages} {165139} (\bibinfo {year}
  {2011})}\BibitemShut {NoStop}%
\bibitem [{SPT()}]{SPTnote}%
  \BibitemOpen
  \href@noop {} {\bibinfo {title} {{We note that while 1D SPT phases are much
  simpler in their structure than phases with intrinsic topological order in 2D
  --- featuring degeneracies due to localized edge modes rather than
  non-trivial underlying topology (i.e. genus $g \neq 0$) --- studying them
  still provides useful insights into the general problem of numerically
  determining many degenerate states.}}}\BibitemShut {Stop}%
\bibitem [{\citenamefont {Fishman}\ \emph {et~al.}(2020)\citenamefont
  {Fishman}, \citenamefont {White},\ and\ \citenamefont
  {Stoudenmire}}]{itensor}%
  \BibitemOpen
  \bibfield  {author} {\bibinfo {author} {\bibfnamefont {M.}~\bibnamefont
  {Fishman}}, \bibinfo {author} {\bibfnamefont {S.~R.}\ \bibnamefont {White}},\
  and\ \bibinfo {author} {\bibfnamefont {E.~M.}\ \bibnamefont {Stoudenmire}},\
  }\href@noop {} {\bibinfo {title} {The \mbox{ITensor} software library for
  tensor network calculations}} (\bibinfo {year} {2020}),\ \Eprint
  {https://arxiv.org/abs/2007.14822} {arXiv:2007.14822} \BibitemShut {NoStop}%
\bibitem [{\citenamefont {Suzuki}(1971)}]{Suzuki1971}%
  \BibitemOpen
  \bibfield  {author} {\bibinfo {author} {\bibfnamefont {M.}~\bibnamefont
  {Suzuki}},\ }\bibfield  {title} {\bibinfo {title} {{Relationship among
  Exactly Soluble Models of Critical Phenomena. I): 2D Ising Model, Dimer
  Problem and the Generalized XY-Model}},\ }\href
  {https://doi.org/10.1143/PTP.46.1337} {\bibfield  {journal} {\bibinfo
  {journal} {Progress of Theoretical Physics}\ }\textbf {\bibinfo {volume}
  {46}},\ \bibinfo {pages} {1337} (\bibinfo {year} {1971})}\BibitemShut
  {NoStop}%
\bibitem [{\citenamefont {Raussendorf}\ and\ \citenamefont
  {Briegel}(2001)}]{Raussendorf2001}%
  \BibitemOpen
  \bibfield  {author} {\bibinfo {author} {\bibfnamefont {R.}~\bibnamefont
  {Raussendorf}}\ and\ \bibinfo {author} {\bibfnamefont {H.~J.}\ \bibnamefont
  {Briegel}},\ }\bibfield  {title} {\bibinfo {title} {{A One-Way Quantum
  Computer}},\ }\href {https://doi.org/10.1103/PhysRevLett.86.5188} {\bibfield
  {journal} {\bibinfo  {journal} {Phys. Rev. Lett.}\ }\textbf {\bibinfo
  {volume} {86}},\ \bibinfo {pages} {5188} (\bibinfo {year}
  {2001})}\BibitemShut {NoStop}%
\bibitem [{\citenamefont {Haldane}(1983)}]{Haldane1983}%
  \BibitemOpen
  \bibfield  {author} {\bibinfo {author} {\bibfnamefont {F.~D.~M.}\
  \bibnamefont {Haldane}},\ }\bibfield  {title} {\bibinfo {title} {{Nonlinear
  Field Theory of Large-Spin Heisenberg Antiferromagnets: Semiclassically
  Quantized Solitons of the One-Dimensional Easy-Axis N\'eel State}},\ }\href
  {https://doi.org/10.1103/PhysRevLett.50.1153} {\bibfield  {journal} {\bibinfo
   {journal} {Phys. Rev. Lett.}\ }\textbf {\bibinfo {volume} {50}},\ \bibinfo
  {pages} {1153} (\bibinfo {year} {1983})}\BibitemShut {NoStop}%
\bibitem [{\citenamefont {Verresen}\ \emph {et~al.}(2017)\citenamefont
  {Verresen}, \citenamefont {Moessner},\ and\ \citenamefont
  {Pollmann}}]{Verresen2017}%
  \BibitemOpen
  \bibfield  {author} {\bibinfo {author} {\bibfnamefont {R.}~\bibnamefont
  {Verresen}}, \bibinfo {author} {\bibfnamefont {R.}~\bibnamefont {Moessner}},\
  and\ \bibinfo {author} {\bibfnamefont {F.}~\bibnamefont {Pollmann}},\
  }\bibfield  {title} {\bibinfo {title} {One-dimensional symmetry protected
  topological phases and their transitions},\ }\href
  {https://doi.org/10.1103/PhysRevB.96.165124} {\bibfield  {journal} {\bibinfo
  {journal} {Phys. Rev. B}\ }\textbf {\bibinfo {volume} {96}},\ \bibinfo
  {pages} {165124} (\bibinfo {year} {2017})}\BibitemShut {NoStop}%
\bibitem [{Note1()}]{Note1}%
  \BibitemOpen
  \bibinfo {note} {In this work and in Ref.~\cite {Roberts2017}, a
  Suzuki-Trotter decomposition is used to efficiently compute an approximation
  to this operator.}\BibitemShut {Stop}%
\bibitem [{Note2()}]{Note2}%
  \BibitemOpen
  \bibinfo {note} {As mentioned in the introduction, we use a DMRG
  implementation provided by \cite {itensor} that performs a 2-site update on
  an initial random product state and include density matrix
  corrections}\BibitemShut {NoStop}%
\bibitem [{\citenamefont {Arad}\ \emph {et~al.}(2017)\citenamefont {Arad},
  \citenamefont {Landau}, \citenamefont {Vazirani},\ and\ \citenamefont
  {Vidick}}]{arad_rigorous_2017}%
  \BibitemOpen
  \bibfield  {author} {\bibinfo {author} {\bibfnamefont {I.}~\bibnamefont
  {Arad}}, \bibinfo {author} {\bibfnamefont {Z.}~\bibnamefont {Landau}},
  \bibinfo {author} {\bibfnamefont {U.}~\bibnamefont {Vazirani}},\ and\
  \bibinfo {author} {\bibfnamefont {T.}~\bibnamefont {Vidick}},\ }\bibfield
  {title} {\bibinfo {title} {Rigorous {RG} {Algorithms} and {Area} {Laws} for
  {Low} {Energy} {Eigenstates} in {1D}},\ }\href
  {https://doi.org/10.1007/s00220-017-2973-z} {\bibfield  {journal} {\bibinfo
  {journal} {Communications in Mathematical Physics}\ }\textbf {\bibinfo
  {volume} {356}},\ \bibinfo {pages} {65} (\bibinfo {year} {2017})}\BibitemShut
  {NoStop}%
\bibitem [{\citenamefont {Affleck}\ \emph {et~al.}(1987)\citenamefont
  {Affleck}, \citenamefont {Kennedy}, \citenamefont {Lieb},\ and\ \citenamefont
  {Tasaki}}]{Affleck1987}%
  \BibitemOpen
  \bibfield  {author} {\bibinfo {author} {\bibfnamefont {I.}~\bibnamefont
  {Affleck}}, \bibinfo {author} {\bibfnamefont {T.}~\bibnamefont {Kennedy}},
  \bibinfo {author} {\bibfnamefont {E.~H.}\ \bibnamefont {Lieb}},\ and\
  \bibinfo {author} {\bibfnamefont {H.}~\bibnamefont {Tasaki}},\ }\bibfield
  {title} {\bibinfo {title} {Rigorous results on valence-bond ground states in
  antiferromagnets},\ }\href {https://doi.org/10.1103/PhysRevLett.59.799}
  {\bibfield  {journal} {\bibinfo  {journal} {Phys. Rev. Lett.}\ }\textbf
  {\bibinfo {volume} {59}},\ \bibinfo {pages} {799} (\bibinfo {year}
  {1987})}\BibitemShut {NoStop}%
\bibitem [{\citenamefont {Jiang}\ \emph {et~al.}(2012)\citenamefont {Jiang},
  \citenamefont {Wang},\ and\ \citenamefont {Balents}}]{Jiang2012}%
  \BibitemOpen
  \bibfield  {author} {\bibinfo {author} {\bibfnamefont {H.-C.}\ \bibnamefont
  {Jiang}}, \bibinfo {author} {\bibfnamefont {Z.}~\bibnamefont {Wang}},\ and\
  \bibinfo {author} {\bibfnamefont {L.}~\bibnamefont {Balents}},\ }\bibfield
  {title} {\bibinfo {title} {Identifying topological order by entanglement
  entropy},\ }\href {https://doi.org/10.1038/nphys2465} {\bibfield  {journal}
  {\bibinfo  {journal} {Nature Physics}\ }\textbf {\bibinfo {volume} {8}},\
  \bibinfo {pages} {902} (\bibinfo {year} {2012})}\BibitemShut {NoStop}%
\bibitem [{\citenamefont {Cincio}\ and\ \citenamefont
  {Vidal}(2013)}]{Cincio2013}%
  \BibitemOpen
  \bibfield  {author} {\bibinfo {author} {\bibfnamefont {L.}~\bibnamefont
  {Cincio}}\ and\ \bibinfo {author} {\bibfnamefont {G.}~\bibnamefont {Vidal}},\
  }\bibfield  {title} {\bibinfo {title} {Characterizing topological order by
  studying the ground states on an infinite cylinder},\ }\href
  {https://doi.org/10.1103/PhysRevLett.110.067208} {\bibfield  {journal}
  {\bibinfo  {journal} {Phys. Rev. Lett.}\ }\textbf {\bibinfo {volume} {110}},\
  \bibinfo {pages} {067208} (\bibinfo {year} {2013})}\BibitemShut {NoStop}%
\bibitem [{\citenamefont {Szasz}\ \emph {et~al.}(2020)\citenamefont {Szasz},
  \citenamefont {Motruk}, \citenamefont {Zaletel},\ and\ \citenamefont
  {Moore}}]{Szasz2020}%
  \BibitemOpen
  \bibfield  {author} {\bibinfo {author} {\bibfnamefont {A.}~\bibnamefont
  {Szasz}}, \bibinfo {author} {\bibfnamefont {J.}~\bibnamefont {Motruk}},
  \bibinfo {author} {\bibfnamefont {M.~P.}\ \bibnamefont {Zaletel}},\ and\
  \bibinfo {author} {\bibfnamefont {J.~E.}\ \bibnamefont {Moore}},\ }\bibfield
  {title} {\bibinfo {title} {Chiral spin liquid phase of the triangular lattice
  hubbard model: A density matrix renormalization group study},\ }\href
  {https://doi.org/10.1103/PhysRevX.10.021042} {\bibfield  {journal} {\bibinfo
  {journal} {Phys. Rev. X}\ }\textbf {\bibinfo {volume} {10}},\ \bibinfo
  {pages} {021042} (\bibinfo {year} {2020})}\BibitemShut {NoStop}%
\bibitem [{\citenamefont {Stoudenmire}\ and\ \citenamefont
  {White}(2013)}]{Stoudenmire2013}%
  \BibitemOpen
  \bibfield  {author} {\bibinfo {author} {\bibfnamefont {E.~M.}\ \bibnamefont
  {Stoudenmire}}\ and\ \bibinfo {author} {\bibfnamefont {S.~R.}\ \bibnamefont
  {White}},\ }\bibfield  {title} {\bibinfo {title} {Real-space parallel density
  matrix renormalization group},\ }\href
  {https://doi.org/10.1103/PhysRevB.87.155137} {\bibfield  {journal} {\bibinfo
  {journal} {Phys. Rev. B}\ }\textbf {\bibinfo {volume} {87}},\ \bibinfo
  {pages} {155137} (\bibinfo {year} {2013})}\BibitemShut {NoStop}%
\bibitem [{\citenamefont {Levy}\ \emph {et~al.}(2020)\citenamefont {Levy},
  \citenamefont {Solomonik},\ and\ \citenamefont {Clark}}]{Levy2020}%
  \BibitemOpen
  \bibfield  {author} {\bibinfo {author} {\bibfnamefont {R.}~\bibnamefont
  {Levy}}, \bibinfo {author} {\bibfnamefont {E.}~\bibnamefont {Solomonik}},\
  and\ \bibinfo {author} {\bibfnamefont {B.~K.}\ \bibnamefont {Clark}},\
  }\href@noop {} {\bibinfo {title} {{Distributed-Memory DMRG via Sparse and
  Dense Parallel Tensor Contractions}}} (\bibinfo {year} {2020}),\ \Eprint
  {https://arxiv.org/abs/2007.05540} {arXiv:2007.05540} \BibitemShut {NoStop}%
\bibitem [{RRG()}]{RRGcode}%
  \BibitemOpen
  \href {https://github.com/brendenroberts/RigorousRG} {\bibinfo {title}
  {{https://github.com/brendenroberts/RigorousRG}}}\BibitemShut {NoStop}%
\bibitem [{\citenamefont {Smolensky}(1986)}]{Smolensky1986}%
  \BibitemOpen
  \bibfield  {author} {\bibinfo {author} {\bibfnamefont {P.}~\bibnamefont
  {Smolensky}},\ }\bibfield  {title} {\bibinfo {title} {{Information Processing
  in Dynamical Systems: Foundations of Harmony Theory}},\ }in\ \href@noop {}
  {\emph {\bibinfo {booktitle} {{Parallel Distributed Processing: Explorations
  in the Microstructure of Cognition}}}},\ Vol.~\bibinfo {volume} {1},\
  \bibinfo {editor} {edited by\ \bibinfo {editor} {\bibfnamefont {D.~E.}\
  \bibnamefont {Rumelhart}}\ and\ \bibinfo {editor} {\bibfnamefont {J.~L.}\
  \bibnamefont {McLelland}}}\ (\bibinfo  {publisher} {MIT press},\ \bibinfo
  {address} {Cambridge},\ \bibinfo {year} {1986})\ Chap.~\bibinfo {chapter}
  {6}, p.\ \bibinfo {pages} {194}\BibitemShut {NoStop}%
\bibitem [{\citenamefont {Carleo}\ and\ \citenamefont
  {Troyer}(2017)}]{Carleo2017}%
  \BibitemOpen
  \bibfield  {author} {\bibinfo {author} {\bibfnamefont {G.}~\bibnamefont
  {Carleo}}\ and\ \bibinfo {author} {\bibfnamefont {M.}~\bibnamefont
  {Troyer}},\ }\bibfield  {title} {\bibinfo {title} {Solving the quantum
  many-body problem with artificial neural networks},\ }\href
  {https://doi.org/10.1126/science.aag2302} {\bibfield  {journal} {\bibinfo
  {journal} {Science}\ }\textbf {\bibinfo {volume} {355}},\ \bibinfo {pages}
  {602} (\bibinfo {year} {2017})}\BibitemShut {NoStop}%
\end{thebibliography}%

\section*{Appendix A: RRG Hyper-parameters and larger system size \label{app:hyper-params}}

\begin{figure}[t]
    \includegraphics[scale=0.5]{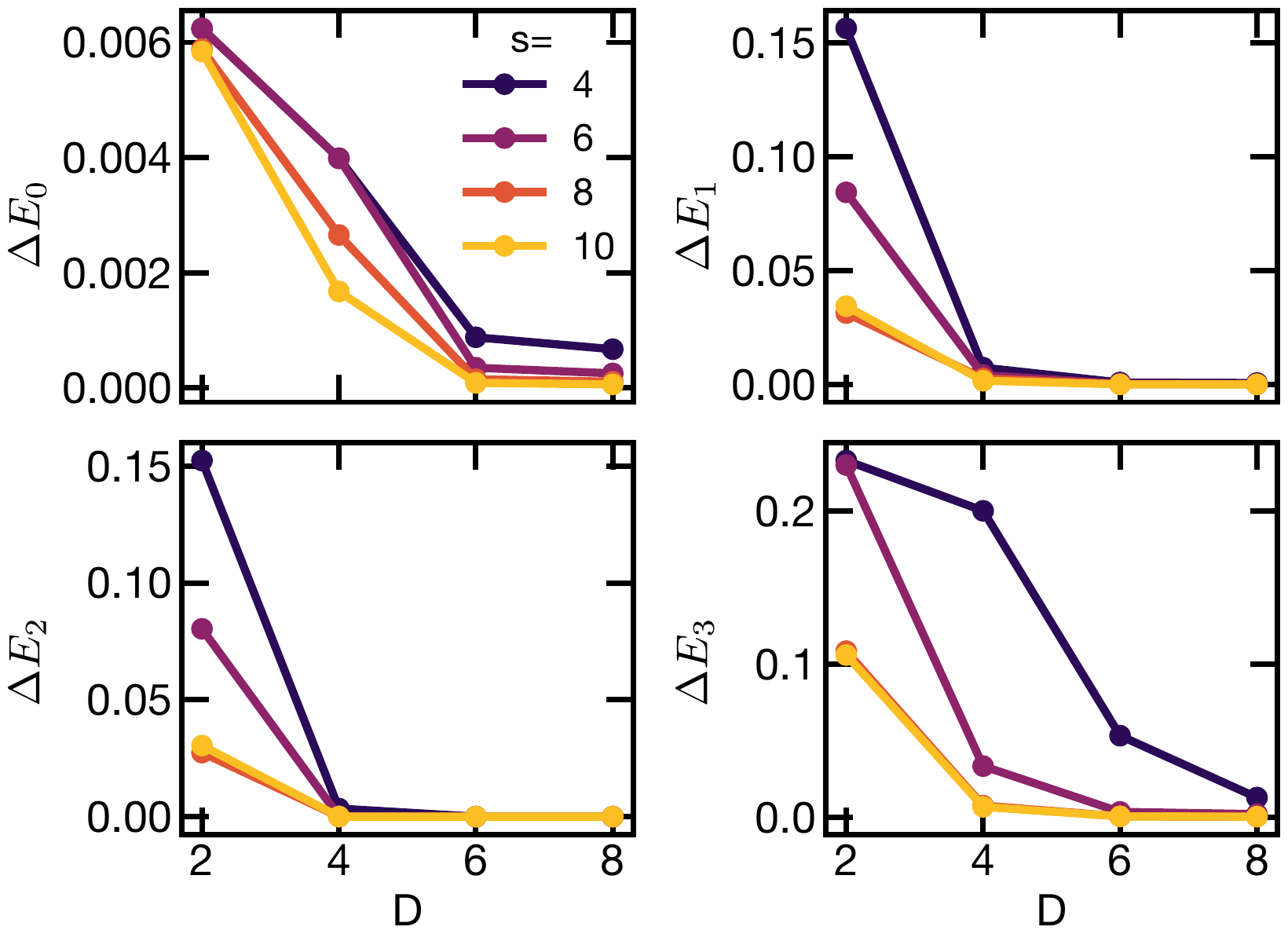}
    \caption{Energy difference between DMRG and RRG at the at critical point of the Tricritical Ising model for varying $s$ (panel) and $D$ ($x$-axis). The ground state energy determined by RRG becomes accurate at low $s$ for sufficiently large $D$.}
    \label{fig:summ-s-D}
\end{figure}

\begin{figure*}
    \includegraphics[scale=0.35]{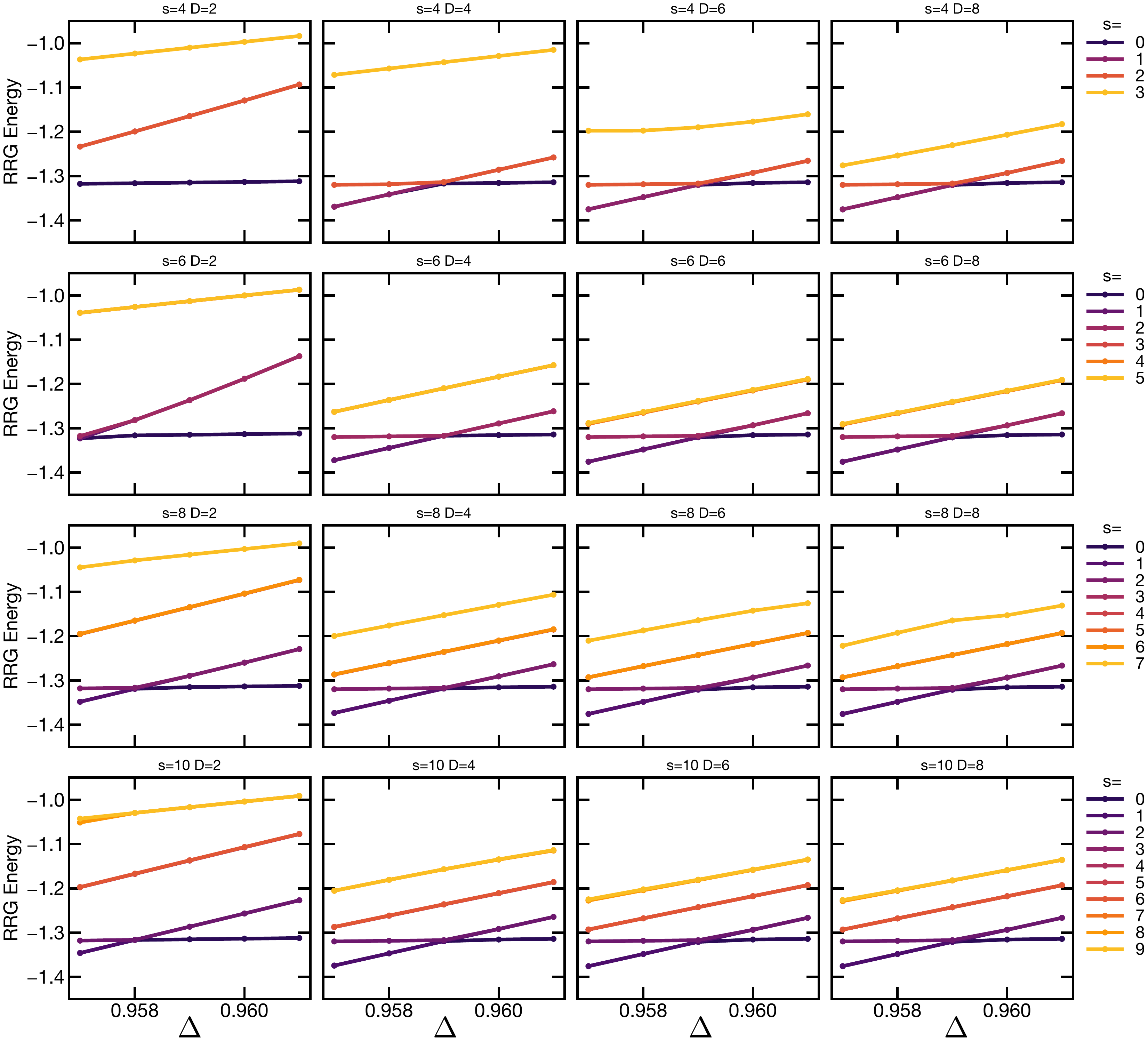}
    \caption{The first $s$ energy levels determined by RRG near the Tricritical Ising model critical point for varying $s,D$. The ground state energy (dark purple) depends on most strongly on $D$, and can be inaccurate even for large $s$ if $D$ is too small (see $s=10, D=2$). Large $s$ and $D$ primarily improve the accuracy of the excited state energies, and may not be required for analyzing the ground state or locating a first order phase transition.}
    \label{fig:table-s-D}
\end{figure*}

Although the focus of our work is developing a qualitative understanding of the advantages of RRG, we have also examined how RRG performance depends on the essential hyper-parameters of the algorithm: the dimension of the targeted subspace, $s$, and the number of local operators used from the AGSP decomposition, $D^2$.
Specifically, we studied the energy discrepancy between the lowest four energy levels determined by RRG and DMRG at the transition point of the ($N=32$) Tricritical Ising model for various $s$ and $D$ (Fig.~\ref{fig:summ-s-D}).
Our results indicate that while increasing either $s$ or $D$ improves RRG accuracy, $D$ has the more pronounced effect, especially on the accuracy of low-lying excited states. 
Indeed, increasing $D=2$ to $D=8$ consistently improves energy accuracy by over an order of magnitude for a wide selection of $s$. 
This behavior can intuitively be explained by recognizing that the AGSP, and hence $D$, controls the renormalization \emph{flow} performed by RRG which is ultimately more important the UV initial conditions, controlled by $s$.
Of course, $s$ should always be at least as large as the degeneracy enforced by symmetry, so that the local Hilbert spaces used to start the renormalization flow will span the entire ground state manifold.
Finally, we note that the energies of lower lying states are generally determined more accurately than those of higher lying states for given $s$ and $D$.
For example, the energy discrepancy to the DMRG value of the ground state at $s=4$, $D=2$ is already $\sim 10^{-3}$, while the discrepancy of the third excited state ($E_3$) is $\sim 10^{-1}$ for the same parameters (Fig.~\ref{fig:summ-s-D}).
We show the full table of the energy dependence on $s$ and $D$ in the region around the transition point of the tricritcal Ising model in Fig.~\ref{fig:table-s-D}.

\begin{figure*}
    \includegraphics[scale=0.5]{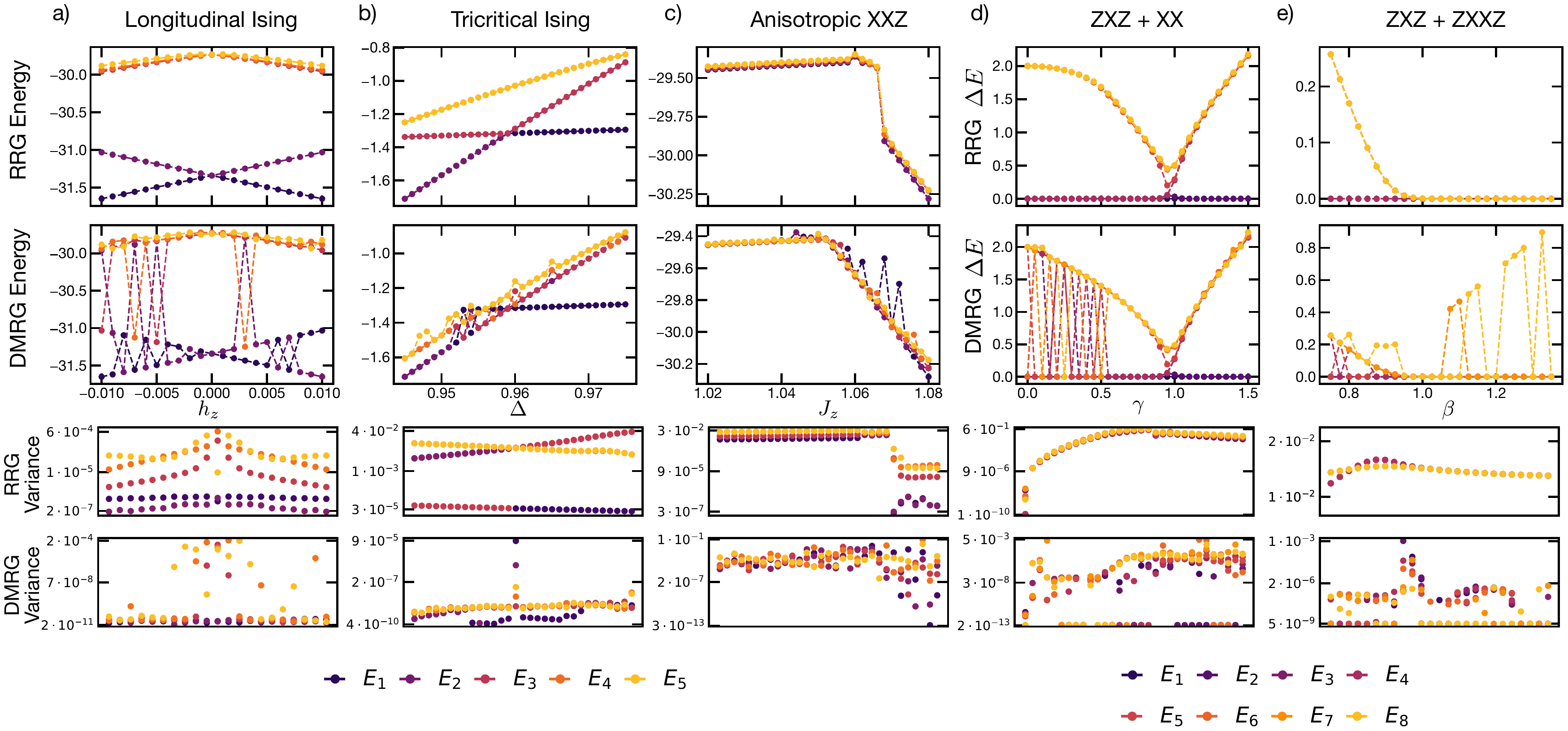}
    \caption{Comparison of energies and variances of the states found using RRG ($D=4$) and DMRG for the different models ($N=32$). (a)~Transverse field Ising model ($s=5$). (b)~Tricritical Ising model ($s=5$). (c)~Anisotropic $XXZ$ model ($s=5$). (d)~$ZXZ + XX$ model ($s=8$). (e)~$ZXZ+ZXXZ$ model ($s=8)$.}
    \label{fig:N32-D4}
\end{figure*}

\begin{figure*}[t]
    \includegraphics[scale=0.5]{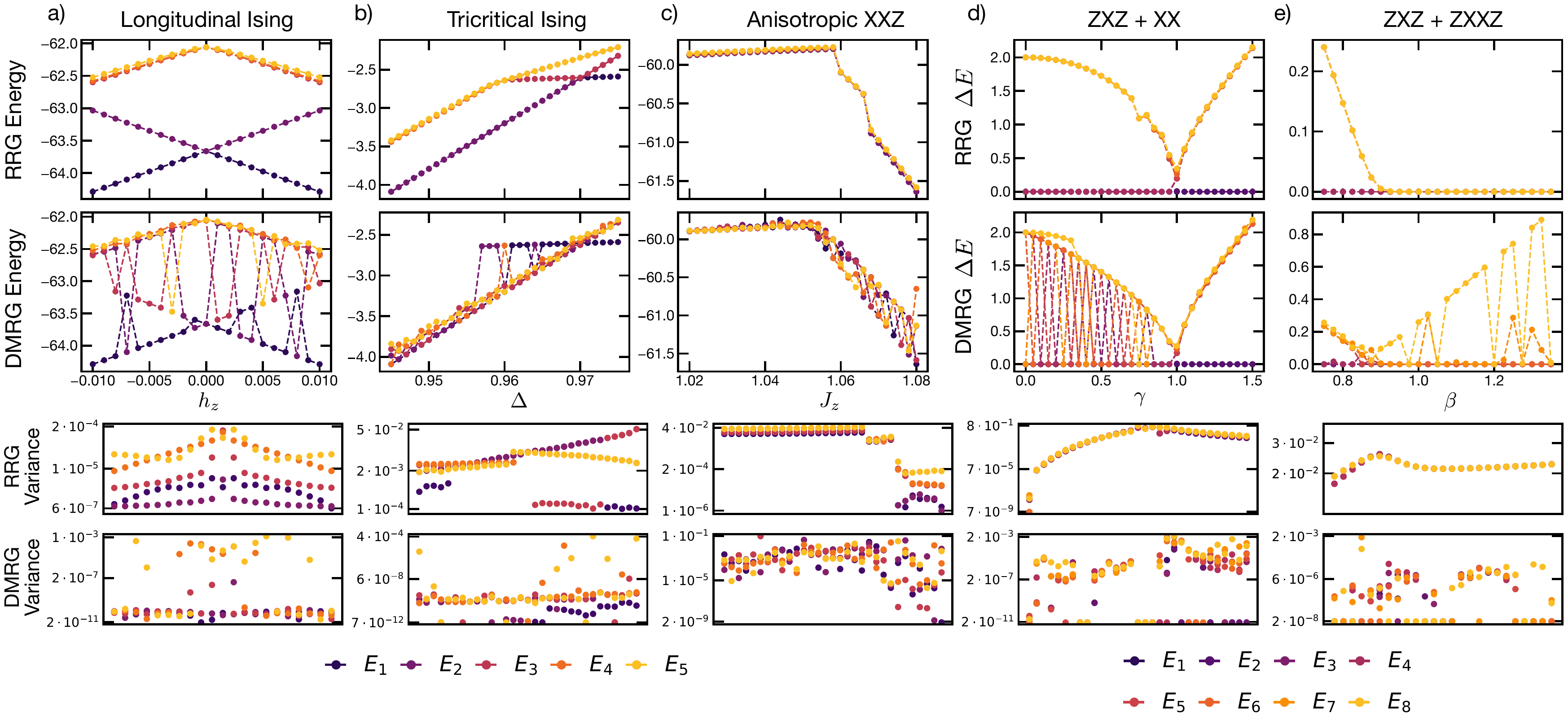}
    \caption{Comparison of energies and variances of the states found using RRG ($D=4$) and DMRG for the different models ($N=64$). (a)~Transverse field Ising model ($s=5$). (b)~Tricritical Ising model ($s=5$). (c)~Anisotropic $XXZ$ model ($s=5$). (d)~$ZXZ + XX$ model ($s=8$). (e)~$ZXZ+ZXXZ$ model ($s=8)$.}
    \label{fig:N64-D4}
\end{figure*}

These findings are confirmed by a broader examination of the effects of reducing $s$ and $D$ across all models.
Fig.~\ref{fig:N32-D4} compares DMRG and RRG ($N=32$) with $D=4$, $s=5$ (magnetic models) or $s=8$ (topological models). 
As expected, for the magnetic models, reducing $D=8$ (used for the results in the main) text to $D=4$ significantly raises the excited state energies determined by RRG, while the lower energy states remain accurate.
This effect is so severe in the tricritical Ising model that RRG actually fails to find a level crossing between $E_3$ and $E_4$, although it determines the $E_1$, $E_2$ crossing --- the phase transition --- correctly.  
The effects of reducing $D$ are also drastic in the anisotropic XXZ model, where it appears RRG now longer accurately determines the phase transition. 
This is in contrast to RRG performance on the topological models, where the effects of reducing $D$ are greatly suppressed.
Indeed, within the topological phases the variance of the RRG states remains relatively small and comparable to those found with $D=8$.
We attribute this to fact that the topological fixed points we consider are frustration free, implying the ground states of the block Hamiltonians can be fused to form the true ground state, without using the AGSP to expand local Hilbert spaces.

Finally, decreasing $s$ and $D$ reduces RRG resource requirements, allowing us to apply RRG to larger system sized ($N=64$).
The accuracy of RRG suffers due to the increased system size, and its troubles with the anisotropic XXZ model and the excited states of the tricritical Ising model are even more pronounced. 
However, aside from the anisotropic XXZ model, RRG still finds an accurate approximation of the ground manifold with proper degeneracy, even in the highly degenerate topological models.
In fact, the ground state variances are not much larger than in the $N=32$ simulations, suggesting that scaling system size is not rapidly compromising RRG's accuracy in the ground state manifold.
This is encouraging, since it may be possible to optimize RRG to be fast enough at low $s$ and $D$ to provide useful seeding to DMRG in future studies.

\section*{Appendix B: Topological Models \label{app:topo-models}}

\begin{figure}[b!]
 \includegraphics[width=\columnwidth]{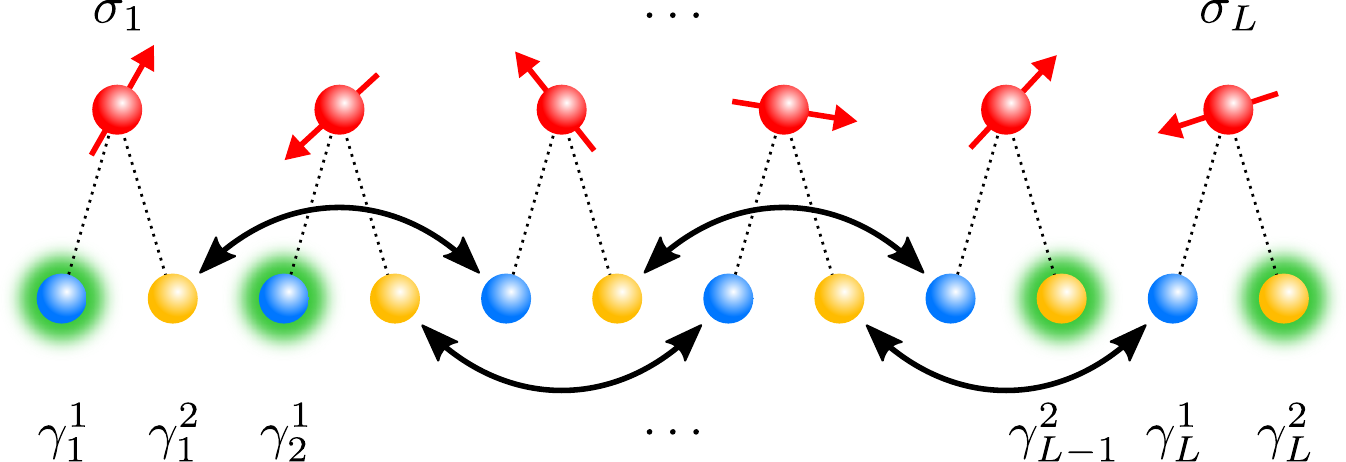}
 \caption{Graphical representation of the $ZXZ$ Hamiltonian. Each spin is replaced by two Majorana modes and the black arrows depict the coupling terms in the Hamiltonian. The modes $\gamma_1^1, \gamma_2^1, \gamma_{L-1}^2$ and $\gamma_L^2$ (highlighted in green) are uncoupled.}
 \label{fig:majo}
\end{figure}

Here, we briefly review the topological models treated in the main text, which include terms of the form
\begin{equation}
        H_{\rm top} = \sum_{j=2}^{L-N} \sigma_{j-1}^z \sigma_j^x \cdots \sigma_{j+N-1}^x \sigma_{j+N}^z \text{ for } N=1,2
\end{equation}
These Hamiltonians can be understood through either the stabilizer formalism or via a Majorana representation~\cite{Verresen2017}.
Let us focus on the most elementary case $N=1$ as it straightforwardly generalizes to larger $N$.
Each spin can be represented by two Majorana fermions with the relation between spin and Majorana operators $\gamma^1_j$ and $\gamma^2_j$ given by
\begin{align}
 \sigma_j^x &= i \gamma^1_j \gamma^2_j, \nonumber \\ 
 \sigma_j^y &= \prod_{k<j} (i \gamma_k^1 \gamma_k^2) \gamma_j^2, \\
 \sigma_j^z &= \prod_{k<j} (i \gamma_k^1 \gamma_k^2) \gamma_j^1. \nonumber
\end{align}
The Hamiltonian of the $N=1$ case with term $\sigma_{i-1}^z \sigma^x_i \sigma^z_{i+1}$ then reads
\begin{equation}
 H = \sum_{j=2}^{L-1} i \gamma^2_{j-1} \gamma^1_{j+1}.
\end{equation}
It becomes obvious from the Hamiltonian and the graphical representation in Fig.~\ref{fig:majo} that two Majorana modes are left uncoupled on either end of the chain.
One can then form localized spin-$1/2$ degrees of freedom from these Majoranas so that the ground state belongs to the same topological class as the one of the AKLT model exhibiting a fourfold ground state degeneracy.
Each increase of $N$ by one adds a free Majorana mode at either edge leading to an additional factor of two in the ground state degeneracy meaning that the ground state of the general model is $2^{N+1}$-fold degenerate.
For even $N$, one Majorana from each edge must pair to form a non-local degree of freedom contributing to the degeneracy which results in symmetry breaking in the spin language~\cite{Verresen2017}.

\renewcommand{\arraystretch}{1.5}
\begin{table*}[t]
    \centering \begin{tabular}{c||c|c||c|c} 
        \hline 
        Model & \multicolumn{2}{c||}{RRG (D=4)} & \multicolumn{2}{c}{DMRG} \\
        \hline \hline
        N=32 & Avg. Time [s] & Avg. Mem. [MB] & Avg. Time [s] & Avg. Mem. [MB] \\
        \hline
        Longitudinal Ising & 19 & 9.7 & 3 & 0.47 \\
        \hline
        Tricritcal Ising & 927 & 98.0 & 19 & 0.72 \\        
        \hline
        Anisotropic XXZ & $9.5 \cdot 10^3$ & 501 & 465 & 3.6 \\
        \hline
        ZXZ + XX & $9.1 \cdot 10^3$ & 471 & $1.8 \cdot 10^3$ & 6.8 \\
        \hline
        ZXZ + ZXXZ & $5.3 \cdot 10^3$ & 265 & 531 & 4.5 \\
        \hline \hline
        N=64 &  &  &   & \\
        \hline
        Longitudinal Ising & 757 & 110 & 6 & 0.53 \\
        \hline
        Tricritcal Ising & $18.4 \cdot 10^3$ & 586 & 38 & 1.1 \\        
        \hline
        Anisotropic XXZ & $120.7 \cdot 10^3$ & $1.2 \cdot 10^3$ & 714 & 5.4 \\
        \hline
        ZXZ + XX & $141.4 \cdot 10^3$ & $2.4 \cdot 10^3$ & $6.9 \cdot 10^3$ & 24.7 \\
        \hline
        ZXZ + ZXXZ & $83.2 \cdot 10^3$ & $1.4 \cdot 10^3$ & $1.4 \cdot 10^3$ & 10.3 \\
        \hline
    \end{tabular}
    \caption{Computational resource comparison of RRG ($D=4$) and DMRG for the investigated models at $N=32,64$. DMRG performance ($N=32$) differs from Table.~\ref{tab:resources} because $s$ was also chosen to be smaller (see Fig.~\ref{fig:N32-D4} and Fig.~\ref{fig:N64-D4}).}
    \label{tab:app-resources}
\end{table*}


\end{document}